\input amstex
\loadeufm
\loadbold
\documentstyle{amsppt}
\NoRunningHeads
\magnification=1200
\pagewidth{32pc}
\pageheight{42pc}
\vcorrection{1.2pc}
\overfullrule=0pt

\define\cq{\Bbb C_q}
\define\Z2{\Bbb Z^{2}}

\define\ua{\boldkey a}
\define\ub{\boldkey b}
\define\uc{\boldkey c}
\define\ud{\boldkey d}
\define\ue{\boldkey e}

\define\fg{\frak g}
\define\gd{\dot{\frak g}}

\define\LL{\Cal  L}
\define\Le{\hat \Cal  L}
\define\Wme{W_m ^{\epsilon}}
\define\W0-1{W_0 ^{-1}}
\define\a{\alpha}
\define\Wmeo{W_{m_1} ^{\epsilon_1}}
\define\Wmet{W_{m_2} ^{\epsilon_2}}
\define\F{\Cal F}
\define\Xme{X_m^{\epsilon}}
\define\X0-1{x_0^{-1}}
\define\Xmeo{X_{m_1}^{\epsilon_1}}
\define\Xmet{X_{m_2}^{\epsilon_2}}
\topmatter
\title   Principal Realization for the extended affine Lie algebra of type
$sl_2$ with coordinates in a simple quantum torus with two generators
\endtitle

\author Stephen Berman and Jacek Szmigielski \endauthor
\affil Department of Mathematics and Statistics\\
University of Saskatchewan, Saskatoon, Canada S7N 5E6 \endaffil
\footnote[]{The authors gratefully acknowledge the support of the Natural
Sciences and Engineering Research Council of Canada }
\dedicatory Dedicated to the memory of Professor Magdi Assem \enddedicatory
\abstract We construct an irreducible representation for the extended affine
algebra of type $sl_2$ with coordinates in a quantum torus.  We explicitly
give formulas using vertex operators similar to those found
in the theory of the infinite rank affine algebra $A_{\infty}$.
\endabstract
\endtopmatter
\document

\subhead {Introduction}\endsubhead
The purpose of this paper is to construct a module for the Extended
Affine Lie algebra (EALA for short), we call it $\Le$, of type $sl_2$
with coordinates in a quantum torus $\cq$.  EALA's were first
introduced in the paper [HKT] as a natural generalization of affine
Kac-Moody algebras and we note that the reported motivation for this
work was from quantum gauge theories.  Roughly one can think of EALA's
as higher dimensional generalizations of loop algebras, and in [HKT]
all examples came from certain central extensions of the Lie algebra
of polynomial maps of an $n$-dimensional torus $T^n$ into a finite
dimensional simple Lie algebra $\gd$ over the field of complex
numbers.  That is, they used Laurent polynomials in many variables as
their coordinates.  Subsequently, in [BGK], it became clear that the
set of axioms for EALA's allowed other coordinate algebras as well.
Indeed, the quantum torus $\cq$ studied in [M], as well as certain
alternative and Jordan algebra coordinates can appear as the
coordinate algebra for an EALA.  In this paper we focus on what is, in
many respects, the easiest example of a coordinate algebra.  Thus, our
coordinates will be a quantum torus, $\cq$, parametrized by one nonzero
scalar $q$ which, in addition,we assume to be generic, in the sense
that it is not a root of unity, and where $\cq$ is generated by two
elements $s
\text{ and } t$ together with their inverses. These generators then satisfy
$$
ts=qst.
$$
Such a quantum torus is
simple as an algebra, and we note it is an algebraic version of the
noncommutative torus studied in [C].  The module we construct is obtained
through a
process which follows the usual construction, by vertex operators, of
the principal module for the affine Kac Moody Lie algebra $A_1 ^{(1)}$ as
found in [LW], [FLM], [K]. Thus, we identify a Heisenberg subalgebra $\hat H$
and use the standard irreducible module for this where the center acts
as the identity to construct our module for $\Le$. Letting $\Cal S= \Bbb C
[x_1,x_2,x_3, \dots ]$ denote this Heisenberg module we will let
$$\F=\Bbb C[v,v^{-1}] \otimes \Cal S,$$ be the space on which we will define
an action of $\Le$ to obtain our module. One cannot fail to notice
that $\F$ is isomorphic to the fermionic Fock space by the
boson-fermion correspondence [K].  This type of fermionic Fock module
arose in the study of the infinite rank affine algebra of type $A_{\infty}$
and the Kadomtsev-Petviashvili
hierarchy [DJKM].  Our formulas are recorded in the bosonic picture, that is
using the symmetric algebra $\Cal S$ rather than the exterior algebra used
for the
fermionic description.  In the bosonic picture the variable $v$
corresponds to the grading
with respect to the variable $t$ of our quantum torus while the
central element of our Heisenberg algebra, $c_s$, is so chosen that it
is tied up with the variable $s$ of our algebra $\Le$.  One may
consult [AABGP] for more on the basics theory of EALA's as well as
[BGK] and [BGKN] for how our algebra $\Le$ fits into this theory. In
[ABGP] one sees how the affine Lie algebras fit into this theory.
However, not very much of the material from these works is needed to
understand the present work, and what is needed will be recalled and
the appropriate references will be cited.  Also, it is clear to us that
the module we construct is very closely related to the basic module
of $A_{\infty}$ [DJKM] and [K].

   Previous to this there has been some work on the representation theory of
toroidal Lie algebras, where we recall these are the universal
 central extensions of Lie algebras of the form
$$\dot {\frak g} \otimes \Bbb C[t_1^{\pm 1},t_2^{\pm 1},\dots ,t_n^{\pm 1}]
\tag 1 $$
where $ \Bbb C[t_1^{\pm 1},t_2^{\pm 1},\dots ,t_n^{\pm 1}]$ is just
the algebra of Laurent polynomials in the commuting variables $t_1,
\dots , t_n$ and $\dot {\frak g}$ is a finite dimensional simple Lie
algebra over $\Bbb C$.  Indeed in [EM] and [EMY1] one finds vertex
operator representations given in the so-called homogeneous (or
untwisted) picture for these toroidal algebras with $\dot {\frak g}$
simply laced while in [Bi1] one finds the principal (or twisted)
picture dealt with.  In these works it is important that one deals with
the universal central extension of the algebra in (1) and also that
the coordinates $ \Bbb C[t_1^{\pm 1},t_2^{\pm 1},\dots ,t_n^{\pm 1}]$
are commutative.  In fact, in these works the authors consider and add
certain derivations, via a generalization of a semi-direct product
construction involving a cocycle,to get a module for a Lie algebra
containing the given toroidal algebra.  However, this bigger algebra is
not necessarily an EALA.  Thus, in the present work one finds, for the
first time, a theory dealing with algebras having noncommutative
coordinates as well as one which gives closed form formulas for the
actions of the derivations involved in our algebras.  Other works
closely related to these are [BC] where one finds the beginnings of a
theory of Verma type modules of some related algebras and the work
[Bi2] where the author constructs an extension of the KdV hierarchy
related to his principal representation given in [Bi1].  The very nice work
[Bi2] is the first to point to definite applications of EALA's to the theory of
integrable systems.  All the works referred to so far are algebraic in
character.  On the analytic side there are several papers discussing
central extensions of current groups, that is $Maps(X;G)$ where $X$ is
a compact manifold and $G$ is a finite dimensional Lie group.  Here we
mention [EF] and [LMNS]. There is also an interesting work [IKUX] in which
a toroidal algebra is identified as the current algebra of
the four-dimensional K\"ahler WZW model.

   In \S1 we recall in detail the definition of our Lie algebra
$\Le$ and the basic properties which we will make use of.  In
particular, we define a particular maximal Heisenberg subalgebra,
$\hat H$, which plays a major role in the rest of the paper.  The very
existence of this Heisenberg subalgebra depends on the noncommutative
nature of our coordinates as it includes certain elements (in the $s$
direction) from the space $[\cq,\cq]I$ which is in our Lie algebra
$\Le$.  Of course in the commutative case, when $q=1$, there are no
such elements available as then the space $[\cq,\cq]$ is zero.  Thus,
it is the very noncommutativity of the coordinates which allows us to
define this Heisenberg subalgebra and hence, in the final analysis, a
space which will eventually become a module for $\Le$.  The rest of
\S1 is devoted to showing that the algebra $\Le$ is, in fact, the
universal central extension of $sl_2(\cq)$. It is becoming more and
more clear (see [Bi1], [Bi2], [EM], [EMY1], [EMY2], [Ya]) that in order to
have a robust representation theory of $sl_2(\cq)$ one needs to study
its universal central extension and vertex representations thereof.
The main support for that comes undoubtedly from the affine theory and
the underlying physical principle that that the central extension is
a result of a $c$-number anomaly brought about by quantization of currents
[GSW, vol.1, page 302].
It is the universal central extension that in principle classifies all such
anomalies.
  We also note here that
in proving that $\Le$ is in fact the universal central extension of
$sl_2(\cq)$ we need to make use of the structure of our quantum torus
$\cq$ as a Jordan algebra.  This is interesting in its own right and
we present complete proofs of the necessary details.

   In \S2 we display the algebra via formal variables.  Here we follow
the method used in [FLM] for the affine Lie algebras and so work with
formal power series with coefficients in one of our Lie algebras.  As
is usual the formulas involve certain delta functions, and although we
list all of the delta function identities we use, we do not offer
proofs for them as most are quite easy and well-known or follow from
well-know ones.  We note here that in doing this one is essentially
giving a very nice basis of the Lie algebra and the basic motivation
for making the right definitions for the power series we consider
comes from both the affine theory as well as our Lemma 2.9.  We
summarize our results in Theorem 2.29, which should be consulted after
reading the basic definitions, by any reader wishing to skip some of
the more computational details of \S2.

   In the final section, \S3, we define our space $\F$ and and action
of our algebra $\Le$ on this space and then proceed to show this
action makes $\F$ into an $\Le$ module.  Here the motivation for
making the correct definitions comes from the fact that we have a
maximal Heisenberg subalgebra in our centrally closed algebra as well
as the usual result, namely Lemma 14.5 in [K], on solutions to certain
eigenvalue equations.  This seems to go back to at least [KKLW].  We
note that in the basic Definition 3.5 there is a scalar factor,
$b(m,\epsilon)$, which appears and that the motivation for what this
should be is provided in the computations of Lemma 3.10.  We have
written the proof of this Lemma in such a way as to display this.  Our
main results are stated at the end of this section and we close by
 making some remarks about the structure of $\F$. Thanks go to Professor
Yun Gao for pointing out some inaccuracies in an eairlier version of this paper.

   This work began as discussions between the two authors and our dear
friend and colleague Prof. Magdi Assem who came to a sudden death
during 1996 while at the Institute for Advanced Studies in Princeton.
There is no doubt that he would have been a co-author to this paper
had he not been taken from us, and it is with the deepest respect that
we dedicate this paper to him.

\subhead {Section 1. Basics on the Lie Algebra}\endsubhead
We will work over the complex field $\Bbb C$ and begin by recalling the
 definition of the quantum torus in two variables over $\Bbb C$. Let $q$ be any
 non-zero complex number and let $I_q$ be the ideal of the group algebra
 of the free group on two generators $s$ and $t$
 over $\Bbb C$  generated by the
relation $ts=qst$. We let $\cq$ denote the factor algebra and as usual we
identify $s$ and $t$ with their images in $\cq$.
$\cq$ is called the quantum torus on the two generators $s$ and $t$.
Thus $\cq$ has a basis consisting
 of monomials $s^{a_1}t^{a_2}$ where $\ua :=(a_1,a_2)$ is in $\Z2$. We will
sometimes
write $m(\ua)$ for the monomial $s^{a_1}t^{a_2}$ so we have that
$$ m(\ua)m(\ub)=q^{a_2b_1}m(\ua +\ub). \tag 1.1 $$
Basic facts about $\cq$ can be found in [BGK] Proposition 2.44 and Remark 2.45.
See [M] for how it is related to other studies.
For our purposes we need to recall that $\cq$ is simple if and only if $q$ is
 not a root of unity, which is the same as saying that the center of $\cq$
is just the scalars $\Bbb C$ in $\cq$.One says that $q$ is generic in this
 case and from
now on we take this as a basic assumption. Thus  we will be working with a
quantum
torus satisfying the following condition.
\proclaim {Assumption \rom{1.2}} $\cq$ is a simple associative
algebra over
$\Bbb C$.
\endproclaim

 Considering $\cq$ as a Lie algebra
under the commutator product we have, since $q$ is generic,
 that the derived algebra, $[\cq,\cq]$,
satisfies
$$ [\cq,\cq]=\oplus_{\ua \in
\Z2\setminus\{0\}}\Bbb {C} m(\ua). \tag 1.3$$
We also have that
$$ \cq=[\cq,\cq] \oplus \Bbb C. \tag 1.4$$
Next, we recall the definition of the map $\varepsilon :\cq \to \Bbb C$.
\proclaim{Definition 1.5}
Define a $\Bbb C$-linear map $\varepsilon :
\cq\to \Bbb C$
by saying that
$$\varepsilon(m(\ua)) =\cases 1 &\text{  if } \ua=0,\\ 0 &\text{  if }
\ua\in\Z2, \ua\neq 0.\endcases
$$ \endproclaim

   This mapping $\varepsilon$ will play a role throughout the paper.
For now, we
note that $\cq$ is a $\Z2$-graded algebra in the obvious way, and hence,
we have
degree derivations $d_s,d_t$ defined by saying
$$ d_s(m(\ua))=a_1m(\ua),\quad d_t(m(\ua))=a_2m(\ua), \tag 1.6$$
where $\ua =(a_1,a_2)$.  We have $\varepsilon(d_s(m(\ua))=0$,  for all
monomials  $m(\ua),\ua \in \Z2$, and similarly for $d_t$ so that,
$$\varepsilon(d_s(x))=0=\varepsilon(d_t(x)),\text{ for all } x\in
\cq.\tag1.7$$ We also remark that as $q$ is generic then $d_s $ and
$d_t$ span the space of outer derivations of $\cq$ when considered as
an associative algebra over $\Bbb C$.  This follows from [BGK]
Remark 2.52.

   We next want to define the Lie algebras which will play a major role in
this paper.  Thus, we let $gl_2(\cq)$ denote the Lie algebra of all $2 \times 2$
matrices with entries in $\cq$.  Of course $gl_2(\cq)$ is
also an associative algebra over $\Bbb C$ and it is the usual commutator product
which we use when considering this as a Lie algebra.  If $A=\pmatrix
x_{1,1} & x_{1,2} \\ x_{2,1} & x_{2,2} \endpmatrix$ is in $gl_2(\cq)$ we define
 $tr(A)$ by,
$$tr(A)=(x_{1,1}+x_{2,2})+[\cq,\cq] \in \frac {\cq}{[\cq,\cq]}.\tag1.8 $$
It is clear that $tr(AB)=tr(BA)$ for any $A,B \in gl_2(\cq) .$
Also note that identifying $\Bbb C \text{ and } \frac {\cq}{[\cq,\cq]}$
we have that $tr(A)=\varepsilon (x_{1,1}+x_{2,2}).$
Moreover we let
$$sl_2(\cq)=\{ A \in gl_2(\cq)|tr(A)=0 \}, \tag1.9$$
so that $sl_2(\cq)$ is a Lie subalgebra of $gl_2(\cq)$.
 As is usual we let
$E=\pmatrix 0 & 1 \\ 0 & 0 \endpmatrix$ ,
$F=\pmatrix 0 & 0 \\ 1 & 0 \endpmatrix $,
$H=\pmatrix 1 & 0 \\ 0 & -1 \endpmatrix$
 and let $I$ denote the $2 \times 2$
 identity matrix. First, we define a new product on $\cq$ by:
$$x \circ y =\frac{xy+yx}{2} \text{ for all } x,y \in \cq. \tag1.10 $$
We note that this product makes $\cq$ into a
Jordan algebra
 with identity over $\Bbb C$.  This will play a role later on when we compute
 central extensions.  We also have that if  $A=\pmatrix
x_{1,1} & x_{1,2} \\ x_{2,1} & x_{2,2} \endpmatrix$, then
$$A=x_{1,2}E+x_{2,1}F+\frac{1}{2} (x_{1,1}-x_{2,2})H+\frac{1}{2}(x_{1,1}
+x_{2,2})I.
\tag1.11 $$
   It follows from this that $sl_2(\cq)$ is of codimension one in $gl_2(\cq)$.
In fact $sl_2(\cq)$ is a simple Lie algebra and is the derived algebra of
$gl_2(\cq)$. Notice also that the algebras $gl_2(\cq)$ and $sl_2(\cq)$
are both graded by $\Bbb Z^3$ in the following way. Elements of the
one-dimensional
space
$$\Bbb C m(\ua)E \text { have degree } (1,a_1,a_2) \in \Bbb Z^3, \tag1.12$$
$$\Bbb C m(\ua)F \text { have degree } (-1,a_1,a_2) \in \Bbb Z^3, \tag1.13$$
$$\Bbb C m(\ua)H \text { have degree } (0,a_1,a_2) \in \Bbb Z^3, \tag1.14$$
for all $\ua =(a_1,a_2) \in \Bbb Z^2$, while elements in
$$\Bbb C m(\ub)I \text { have degree } (0,b_1,b_2) \in \Bbb Z^3, \tag1.15$$
for all elements $\ub =(b_1,b_2) \in \Bbb Z^2 \setminus \{ 0 \}.$ Thus, letting
$$\Delta = \{(n,a_1,a_2) \in \Bbb Z^3 | n \in \{-1,0,1 \},\ua =(a_1,a_2)
\in \Bbb Z^2 \}
 ,\tag1.16$$ we have that
$$\boldkey n=(n_1,n_2,n_3) \in \Bbb Z^3 \text{ is a degree of a non-zero space
in $sl_2(\cq)$
if and only if } \boldkey n \in \Delta. \tag1.17$$

Similar remarks apply to the algebra $gl_2(\cq)$ and in particular, this is
graded by $\Bbb Z^3$ both as a Lie algebra as well as an associative algebra.
It follows that the derivations $d_s,d_t$ of $\cq$ have natural lifts to degree
 derivations of $gl_2(\cq)$, and we again denote these by $d_s,d_t$.

   Note that we have a bilinear form defined on $gl_2(\cq)$ as follows. For
$X,Y \in gl_2(\cq)$ we let
$$(X,Y)=tr(XY) \in \frac {\cq}{[\cq,\cq]}\cong\Bbb C.  \tag1.18$$
Here, as before, we have identified $\frac {\cq}{[\cq,\cq]} \text{ with }
\Bbb C$.  It is
clear that this defines a symmetric bilinear form which is invariant in the
sense
that for $X,Y,Z \in gl_2(\cq)$ we have
$$([X,Y],Z)=(X,[Y,Z]). \tag1.19 $$ Moreover, the restriction of this
form to $sl_2(\cq)$ is non-degenerate.

Restricting the derivations $d_s,d_t$
 to $sl_2(\cq)$ lets us define a central extension,which we denote as
$\LL$. As a vector space over $\Bbb C$, $\LL$ is given by
$$
\LL=sl_2(\cq) \oplus \Bbb C c_s \oplus \Bbb C c_t, \tag1.20
$$
where $c_s,c_t$ are two new symbols.  Multiplication in $\LL$ is given as
follows.  Let $X,Y \in sl_2(\cq)$, then as elements in $\LL$ we define
$$[X,Y]=XY-YX+(d_s X,Y)c_s+(d_tX,Y)c_t,\tag1.21$$
$$ [c_s,\LL]=\{ 0 \} = [c_t,\LL]. \tag1.22$$
That this defines a Lie algebra is clear from the fact that both $d_s
\text{ and }
d_t$ are skew-symmetric relative to the above bilinear form on
$sl_2(\cq)$.  Thus, the subspace $sl_2(\cq)$ of $\LL$ has co-dimension
2.  We will soon see that, in fact, $\LL$ is the universal central
extension of $sl_2(\cq)$.

   We can now extend the Lie algebra $\LL$ to a Lie algebra $\Le$ by adding the
derivations $d_s,d_t$ to $\LL$. Formally, we let
$$\Le=\LL \oplus \Bbb C d_s \oplus \Bbb C d_t , \tag1.23$$ with multiplication
extending (1.21) and (1.22) and satisfying
$$[d,X]=d(X),[d,c]=0,\text{ for } X \in sl_2(\cq),d \in \{d_s,d_t\},
c \in \{c_s,c_t \}. \tag1.24$$
Letting the elements $c_s,c_t,d_s,d_t$ have degree $(0,0,0) \in \Bbb Z^3$ we see
that both of our algebras have $\Bbb Z^3$ gradings with the degrees of
 the non-zero spaces being in $\Delta$ as in 1.16.  Thus, we can write
$$\LL=\oplus_{\boldkey n \in \Delta} \LL_{\boldkey n} \text{ and }
\Le=\oplus_{\boldkey n \in \Delta} \hat L_{\boldkey n}. \tag1.25$$
Moreover, we have that
$$dim\LL_{\boldkey n}=dim\Le_{\boldkey n}=1 \text{ if }
\boldkey n=( \pm 1,a_1,a_2), \tag1.26$$
$$dim\LL_{\boldkey n}=dim\Le_{\boldkey n}=2 \text{ if }
\boldkey n=( 0,a_1,a_2),\text { and } \ua=(a_1,a_2) \neq 0, \tag1.27$$
$$dim\LL_{\boldkey 0}=3\text{ and }dim\Le_{\boldkey 0}=5. \tag1.28$$
Notice also that $\LL$ is the derived algebra of $\Le$. We extend the form on
$sl_2(\cq)$ to a symmetric form on all of $\Le$ by setting
$$(c_s,c_t)=(d_s,d_t)=0=(c_s,d_t)=(c_t,d_s) \text{ and }
(c_s,d_s)=(c_t,d_t)=1, \tag1.29$$
and then declaring that the subspace spanned by $c_s,c_t,d_s,d_t$ is
perpendicular
to $sl_2(\cq)$.  That this form is a non-degenerate symmetric bilinear form
on $\Le$ which is invariant is easy to check.  In fact,$\Le$ is an
extended affine Lie algebra of type $A_1$, in the sense of [AABGP],
 whose core is just $\LL$.

\remark{Remark \rom{1.30}}In the above we have used the Lie algebra $sl_2(\cq)$
to build the algebras $\LL \text{ and } \Le$ but we could have very well
used $gl_2(\cq)$ in it's place.  At several points in what follows it will be
convenient to work in these bigger algebras obtained by "adding the identity
 as a central element".
\endremark

   Our next goal is to introduce a Heisenberg subalgebra of $\Le$ which
 will play a major role in the rest of the paper.  It is here that the
non-commutative
 nature of the coordinates $\cq$ first makes itself felt as we need to use some
elements from the space $[\cq,\cq]=\oplus_{\ua \in
\Z2\setminus\{0\}}\Bbb {C} m(\ua)$, which is non-trivial in our case.  The
basis
elements which we need are given in the following definition.
\proclaim{Definition 1.31} Define elements $E_{2k+1},A_m \text{ for }
k\in
\Bbb Z
,m\in \Bbb Z \setminus \{ 0 \}$ by saying
$$E_{2k+1}=s^k E+s^{k+1} F \text{ and } A_m=s^m I$$
Furthermore we let $\hat H$ be the span of all of these elements together
with the element $c_s$.
 \endproclaim
\proclaim{Lemma 1.32} The subspace $\hat H$ is an infinite dimensional
Heisenberg
Lie algebra which is maximal in the sense that it is not properly contained in
any other Heisenberg subalgebra of $\Le$.
\endproclaim

\demo{Proof}It is easy to check that the following formulas hold.
$$[E_{2k+1},E_{2j+1}]=(2k+1)\delta_{k+j+1,0} c_s,\tag1.33$$
$$[A_m,A_n]=2m\delta_{m+n,0} c_s,\tag1.34$$
$$[E_{2k+1},A_m]=0,\text{ for all } k,j \in \Bbb Z \text{ and }
m,n \in \Bbb Z \setminus \{ 0 \}.\tag1.35$$
>From these it is clear that $\hat H$ is a Heisenberg subalgebra of $\Le$.
To see this is maximal, first note that in the Lie algebra $g
l_2(\cq)$
the maximal commuting Lie subalgebra containing the element $E_1$ above
consists of elements of the form $f_1(s)I+f_2(s)E_1 $ where both $
f_1(s) \text { and } f_2(s)$ are just Laurent polynomials of the form
$a_j s^j+ \dots + a_k s^k \text{ for } j \leq k \text{ integers }.$  It now
easily follows from this that $\hat H$ is maximal.
 \qed \enddemo

   Our next goal is to show that $\LL$ is the universal central extension
(uce for short) of the Lie algebra $sl_2(\cq)$.  From knowing the situation
 described in [BGK] one would certainly guess that this is the case. Indeed,
in that work the uce of $sl_n(\cq)$ is described for $n \geq 3$ and any
quantum torus in any number of variables
using the results from [KL]. See also [BGKN] for related material.
In these cases it turns out that the Connes first cyclic homology
group, $HC_1(\cq)$ is two dimensional when the quantum torus, $\cq$, is simple
and generated by two elements and their inverses, as in the case we are
considering
 in this paper.  This homology group, $HC_1(\cq)$, is defined using the
associative algebra structure of $\cq$ while in the case we are interested
in here, namely with $sl_2(\cq)$, it is the Jordan algebra structure,under
the Jordan product $\circ$ of
(1.10), which
 comes into play.  Roughly speaking, $sl_2(\cq)$ is too small so that  it only
"sees" the Jordan algebra structure of the coordinates and not the associative
algebra structure. The reader should see [AG],[ABG] and [G] for related
material.  Note
also that the paper [Y] leads one to expect that $HC_1(\cq)$ (so considering
$\cq$ as an associative algebra) will not give this uce when q is chosen to
be an even root of unity.

   There are various methods we could use here and we have chosen the one that
closely resembles that used in [BGK] and goes back to Tits [T].  Thus we
closely follow the
 presentation in [Se], pages 61-65 and we only indicate the argument until
we come to the end where we must compute using the Jordan product on $\cq$.

   To begin, we let $\fg$ be the uce of $sl_2(\cq)$ which exists as this latter
algebra is perfect.  Then we have a surjective homomorphism $$\phi :\fg
\rightarrow
sl_2(\cq).  \tag1.36 $$ Pulling back the subalgebra of $sl_2(\cq)$ generated
by our three elements $E,F,H$ we can get (after normalizing by a central
element)
an $sl_2 -triple$ of elements in $\fg$.  Then as in [Se] one decomposes $\fg$
into a
direct sum of $sl_2(\Bbb C)$ modules each of which is either isomorphic to
 the trivial one
dimensional module or to the adjoint module.  We thus obtain,
$$\fg=\Cal D \oplus (sl_2(\Bbb C) \otimes \cq), \tag1.37$$
where there is a product $\{ \cdot ,\cdot \} :\cq \times \cq \rightarrow
\Cal D$ as well as an action of $\Cal D \text{ on } \cq$ by
derivations when $\cq$ is considered as a Jordan algebra under $\circ$.
One has $\Cal D = \{ \cq, \cq \}$ and  with proper normalizations
the following formulas hold,
$$[x \otimes a, y \otimes b]=B(x,y) \{ a,b \} +[x,y] \otimes a \circ b,
\tag1.38 $$
$$[ \{ a,b \},x \otimes c]= x \otimes (\{ a,b \}(c)), \tag 1.39$$
$$[ \{ a,b \}, \{ c,d \} ]=\{ \{ a,b \}(c), d \}+ \{c, \{ a,b \} (d) \} ,
\tag1.40$$
where $ x,y \in sl_2(\Bbb C),a,b,c,d \in \cq \text{ and where } B( \cdot,
\cdot)$
is the Killing form on $sl_2(\Bbb C)$.  In the above we have
$$\{ a,b \} (c) = \frac{1}{2}(a \circ (b \circ c) -b \circ (a \circ c))=
\frac{1}{8}[[a,b],c], \tag1.41 $$
$$ \{ a,b \}=- \{ b,a \},\text{ and } \tag 1.42$$
$$\{ a \circ b, c\} +\{ b \circ c, a \} +\{ c \circ a, b \}=0. \tag 1.43$$
Thus, the homomorphism $\phi$ maps $x \otimes a \text{ to } ax \in sl_2(\cq)
\text { for } x \in \{ E,F,H \}.$  We also have that $\phi (\{ a,b \})=
\frac{1}{8} [a,b]I \in \cq.$

   The next step is to define a universal object satisfying the conditions
 satisfied by $\Cal D$ above, so conditions 1.42 and 1.43. Thus, we let
$$ \langle \cq,\cq \rangle =\frac{\cq \otimes \cq }{T}, \tag 1.44$$
 where $T$ is the subspace of $\cq \otimes \cq$ spanned by the
following elements for $a,b,c \in \cq$
$$ a \otimes b +b \otimes a, a \circ b \otimes c+b \circ c \otimes a +
c \circ a+b \text{ for } a,b,c \in \cq. \tag 1.45 $$
Notice that here we have to use the Jordan product and this is what
distinguishes
this case from that considered in [BGK].  Using the object $\langle \cq,\cq
\rangle$
we form the space $\hat {\fg}$ defined by
$$\hat {\fg}=\langle \cq,\cq \rangle \oplus (sl_2(\Bbb C) \otimes \cq)
\tag1.46$$
and make this into an algebra by using formulas 1.38, 1.39, 1.40, and 1.41
but with $\langle \cdot,\cdot \rangle \text{ replacing }
\{ \cdot,\cdot \}.$ It is straightforward to verify this is in fact a
 well-defined Lie
algebra which is a perfect central extension of $sl_2(\cq)$.  Thus we get a Lie
 algebra homomorphism $\mu \text{ from } \fg \text { onto } \hat{\fg}.$
But using
the universal property inherent in our definition of $\langle \cq,\cq \rangle$
together with the definition of multiplication in $\hat {\fg}$ we get a Lie
algebra
 homomorphism $\nu \text{ from } \hat {\fg} \text { onto } \fg.$  As the
compositions
$\nu \circ \mu \text{ and } \mu \circ \nu$ are both identity maps we get that
$\fg \text{ and } \hat {\fg}$ are isomorphic.  Thus, $\hat \fg$ is the uce of
$sl_2(\cq).$

   From what we have so far we know, as $\LL$ is a perfect central extension of
$sl_2(\cq)$, that there is a surjective Lie algebra homomorphism
$\psi :\hat \fg \rightarrow \LL .$ Noting that in $\LL$ we have
$$[m(\ua)E,m(\ub)F]=(m(\ua) \circ m(\ub))H+ \frac{1}{2}[m(\ua),m(\ub)]I+
\delta_{\ua +\ub ,\boldkey 0}q^{-a_1a_2}(a_1c_s+a_2c_t),\text { and } \tag1.47$$
$$[m(\ua)H,m(\ub)H]=[m(\ua),m(\ub)]I+2\delta_{\ua +\ub ,\boldkey 0}q^{-a_1a_2}
(a_1c_s+a_2c_t) \text{ for}\quad  \ua,\ub \in \Bbb Z^2,
 \tag1.48$$
we have that the homomorphism $\psi$ can be taken to satisfy the following
  formulas,
$$\psi (X \otimes m(\ua))=m(\ua)X,\text{ for } X \in \{ E,F,H \} ,\tag 1.49$$
and
$$ \psi (\langle m(\ua),m(\ub) \rangle)=\frac{1}{8}[m(\ua),m(\ub)]I
+\frac{1}{4} \delta_{\ua +\ub ,\boldkey 0}q^{-a_1a_2}
(a_1c_s+a_2c_t) . \tag1.50$$
Also, it is clear that $ \langle \cq, \cq \rangle$ is graded by $\Bbb Z^2$
where the degree of $\langle m(\ua),m(\ub) \rangle$ is just $\ua +\ub.$  Thus,
we have that $\psi$ induces a graded homomorphism between the graded vector
 spaces  $ \langle \cq, \cq \rangle \text{ and } [\cq,\cq] \oplus
\Bbb C c_s \oplus \Bbb C c_t.$
Here we have given $c_s,c_t$ degree $\boldkey 0$.  Writing
$ \langle \cq, \cq \rangle_{\ua} $
for the subspace of$ \langle \cq, \cq \rangle$ of degree $\ua$ we find,
because of 1.3, that
$$dim  \langle \cq, \cq \rangle_{\ua} \geq 1 \text{ if } \ua \neq \boldkey 0
\tag1.51$$
$$dim \langle \cq, \cq \rangle_{\boldkey 0} \geq 2.\tag1.52$$ Thus, in
order to show that $\LL$ is the uce of $sl_2(\cq)$ we need only show
the two inequalities (1.51) and (1.52) are in fact equalities.  We do
this in our next results but need to develop a little more notation
first.

   For $\ua,\ub \in \Bbb Z^2$ we let
$$\sigma (\ua,\ub)=q^{-a_2b_1},f(\ua,\ub)=\sigma (\ua,\ub) \sigma
(\ub,\ua)^{-1},
\text{ and } 2 \rho (\ua,\ub)=\sigma (\ua,\ub) (1+f(\ub,\ua)). \tag1.53$$
Then we have the following formulas for $\ua,\ub \in \Bbb Z^2.$
$$m(\ua)m(\ub)=\sigma (\ua,\ub)m(\ua+\ub),\tag 1.54$$
$$m(\ua)m(\ub)=f(\ua,\ub)m(\ub)m(\ua), \tag1.55$$
$$ m(\ua) \circ m(\ub) =\rho (\ua,\ub) m(\ua+\ub). \tag1.56$$
Moreover, we have $\rho (\ua,\ub) =\rho (\ub,\ua), \text{ and as $q$ is
not an even root of unity, } \rho (\ua,\ub) \neq 0.$

   Notice that we have for any $x,y,z \in \cq$ that
$$\langle x \circ y ,z \rangle +\langle y \circ z, x \rangle +
\langle z \circ x,y \rangle =0, \tag 1.57$$
and that $\langle \cdot ,\cdot \rangle$ is skew-symmetric. This gives us
for any $\ub ,\uc ,\ud \in \Bbb Z^2$, that
$$\rho (\ub,\uc) \langle m( \ub + \uc),m(\ud) \rangle =
\rho (\uc,\ud ) \langle m( \ub),m( \uc + \ud) \rangle +
\rho (\ub,\ud) \langle m(\uc), m(\ub + \ud ) \rangle . \tag1.58 $$
Clearly, by (1.57) we have $\langle 1,x \rangle =0 \text{ and }
\langle x,x \rangle = 0 \text{ for any } x \in \cq.$

\proclaim{Lemma 1.59}For $n \in \Bbb Z,n \neq -1$,we have that,
$\langle s^n,s \rangle=\langle t^n,t \rangle=0.$
\endproclaim

\demo{Proof}For $a,b \in \Bbb Z$ we have that $s^a \text{ and } s^b$
commute so that $s^a \circ s^b=s^{a+b}.$ Thus from (1.57) we have that
$\langle s^a,s^b \rangle=\langle s,s^{a+b-1} \rangle +
\langle s^{a-1},s^{b+1} \rangle.$ Assuming $n \geq 1$ we obtain
 $\langle s^{n+1},s \rangle=\langle s^n \circ s,s \rangle=
-\langle s^{n+1},s \rangle +\langle s^n,s^2 \rangle$. Doing this again to the
second term we get that $\langle s^{n+1},s \rangle=
-2\langle s^{n+1},s \rangle +\langle s^{n-1},s^3 \rangle$.
Proceeding in this
 way after $n$ steps we get that
$\langle s^{n+1},s \rangle=
-n\langle s^{n+1},s \rangle +\langle s^{n-(n-1)},s^{n+1} \rangle$.
But this equals $ -(n+1)\langle s^{n+1},s \rangle$ and so we get that
$\langle s^{n+1},s \rangle =0$.

   For the other case note that $\langle 1,s \rangle =0,$ while we
know that $\langle s^{-1},s \rangle \neq 0$ since it's image under
$\psi \text{ is } - \frac{1}{4} c_s.$  For $n \geq 2$ we have that
$\langle s^{-n},s \rangle=\langle s^{-(n-1)} \circ s^{-1},s \rangle=
-\langle s^{-n+2},s^{-1} \rangle.$ If $n=2$ this is zero.Otherwise, this is
$-\langle s^{-n+3} \circ s^{-1},s^{-1} \rangle=
\langle s^{-n+2},s^{-1} \rangle -\langle s^{-n+3},s^{-2} \rangle.$
Repeating this process $n-2$   times yields that
$\langle s^{-n},s \rangle=-\langle s^{-n+2},s^{-1} \rangle=
(n-2)\langle s^{-n+2},s^{-1} \rangle.$ It follows that
$\langle s^{-n},s \rangle=0$.  The assertion with $t$ replacing $s$ is
proved in the same way. \qed
 \enddemo
We now let $\epsilon_1=(1,0) \text{ and } \epsilon_2 =(0,1) \text{ in }
\Bbb Z^2.$
\proclaim{Lemma 1.60} For all $n \in \Bbb Z,\ue \in \Bbb Z^2$ we have

\roster
\item $(\langle s^n,m(\ue) \rangle \in \Bbb C \langle s,
m(\ue +(n-1) \epsilon_1\rangle$,
\item $\langle t^n,m(\ue) \rangle \in \Bbb C \langle t, m(\ue +(n-1) \epsilon_2
\rangle$,
\endroster
In particular, $dim \langle \cq,\cq \rangle_{\ue} \leq 2, \text{ for all }
\ue \in \Bbb Z^2.$
\endproclaim

\demo{Proof} We only need to prove (1) as (2) is similar.  For $n=0$
the element above is zero while when $n=1$ the result is clear. For $n \geq 2$
we use induction on $n$ and have that
$\langle s^n,m(\ue) \rangle=
\langle s^{n-1} \circ s, m(\ue) \rangle=
\langle s^{n-1},s \circ m(\ue) \rangle + \langle s, m(\ue) \circ s^{n-1}
\rangle.$
Clearly we have that $\langle s, m(\ue) \circ s^{n-1} \rangle \in
\Bbb C \langle s, m(\ue +(n-1) \epsilon_1
\rangle,$ while by induction we get that
$\langle s^{n-1},s \circ m(\ue) \rangle \in
\Bbb C \langle s, m(\ue +(n-1) \epsilon_1
\rangle.$ In a similiar way we have , for $n \geq 1$, that
$\langle s^{-n},m(\ue) \rangle \in
\Bbb C \langle s^{-1}, m(e-(n-1) \epsilon_1) \rangle.$
Finally, note that for any $\ud \in \Bbb Z^2$ we have, by (1.58), that
$\rho(\ud -\epsilon_1,\epsilon_1)\langle m(\ud),s^{-1} \rangle=
\rho (\ud -\epsilon_1,\epsilon_1)\langle m(\ud -\epsilon_1 + \epsilon_1),
s^{-1} \rangle=
\rho (\epsilon_1,-\epsilon_1) \langle m(\ud - \epsilon_1), m(\boldkey 0)
\rangle +
\rho ( \ud-\epsilon_1, -\epsilon_1) \langle s, m(\ud -2 \epsilon_1) \rangle
\in \Bbb C \langle s, m( \ud-2 \epsilon_1) \rangle.$ This is what we want.
The last statement of the Lemma follows from the first two together with the
fact that if $\ua=(a_1,a_2) \in \Bbb Z^2 \text{ then }
\rho (a_1\epsilon_1, a_2\epsilon_2) \langle m(a_1 \epsilon_1 + a_2 \epsilon_2),
m( \ub) \rangle=
\rho (\ub, a_2 \epsilon_2) \langle m(a_1 \epsilon_1), m( \ub +a_2
\epsilon_2) \rangle
+ \rho (\ub, a_1 \epsilon_1) \langle m(a_2 \epsilon_2), m( \ub +a_1
\epsilon_1) \rangle.$
\qed
\enddemo

   We can now complete the proof of our result on universal central extensions.

\proclaim{Theorem 1.61} $\LL$ is the universal central extension of the Lie
algebra $sl_2(\cq)$.
\endproclaim

\demo{Proof} We know that we only need to show equality holds in (1.51) and
(1.52). However we already know from the previous two Lemmas that
$dim \langle \cq,\cq \rangle_{\boldkey 0} =2$, and that if
$\ua=(a_1,a_2)$
with exactly one of $a_1,a_2$ equal to $0$ then
$dim \langle \cq,\cq \rangle_{\ua}$ equals $1$. Finally we suppose that
$\ua =(a_1,a_2) \text{ where both } a_1, a_2 $ are non-zero. Then we have,
by (1.58) that
$\langle m( \ua - \epsilon_2),t \rangle=
\langle m(a_1 \epsilon_1 +(a_2 -1) \epsilon_2),t \rangle=
\alpha \langle m(a_1 \epsilon_1), m(a_2 \epsilon_2) \rangle
+\beta \langle m((a_2 -1) \epsilon_2),m( a_1 \epsilon_1 + \epsilon_2) \rangle,$
for some non-zero scalars $\alpha,\beta.$ Since $q$ is generic and $a_1,a_2$
are non-zero we have that $[m(a_1 \epsilon_1),m(a_2 \epsilon_2)] \neq 0,$
so that $\langle m(a_1 \epsilon_1),m(a_2 \epsilon_2) \rangle \neq 0.$ Thus
we obtain
that the element
 $\alpha \langle m(a_1 \epsilon_1),m(a_2 \epsilon_2) \rangle  =
\langle m( \ua - \epsilon_2),t \rangle - \beta
 \langle m((a_2 -1) \epsilon_2),m( a_1 \epsilon_1 + \epsilon_2) \rangle$
is non-zero and so, by the previous Lemma in the space
$\Bbb C \langle t, m(\ua - \epsilon_2) \rangle \cap
\Bbb C \langle s, \ua -\epsilon_1) \rangle.$ It follows that
$\langle \cq,\cq \rangle_{\ua}$ is one dimensional.
\qed
\enddemo

\subhead {Section 2. Displaying the algebra via formal variables}\endsubhead
In this section we will give what amounts to a, particularly nice,
multiplication table for our Lie algebra $\LL$ using the approach via
formal variables as in [FLM].  As the basic set up is very well known
in the affine case we will be brief and assume the basics on delta
functions and the identities they satisfy.  Thus, as has become
customary we will only state the ones we use and leave their
verification to the reader who should consult [FLM].

   To begin with we note that we will be working with  spaces
of the form $V [[z,z^{-1}]]$, where $V$ is a vector space over $\Bbb C$ and
where
$$
V [[z,z^{-1}]]= \{ \Sigma_{n \in \Bbb Z} v_n z^{-n} |v_n \in  V \}. \tag2.1$$
Often for us in this section the space $V$ will be $\LL \text{ or } \Le$.
However in a few places we use the versions  coming from $gl_2(\cq)$ as
mentioned
in Remark1.30.  Later $V$ will be a space of the form $End( \Cal F)$ for
an appropriately chosen space $\Cal F$. Recall also that the delta
function, $\delta (z)$ is defined by
$$ \delta(z)= \Sigma _{n \in \Bbb Z} z^n. \tag2.2$$
We will also have  need to use $z \frac{d}{dz} \delta (z)$ which we
denote by $\delta ^{(1)} (z).$  Thus,
$$\delta^{(1)} (z) = \Sigma_{n \in \Bbb Z} nz^n. \tag2.3$$

   We define the following elements,
$$e(k,m) :=s^k t^mE,\quad f(k,m) := s^k t^m F, \text{ and } h(k,m) := s^k t^mH
\text{ for } k,m \in \Bbb Z. \tag 2.4$$

One should notice that these elements, together with the elements
$s^k t^m I, c_s, c_t \text { for } k,m \in \Bbb Z \text {, and not both zero, }
$ form a basis of $\LL$.  The  series that shall concern us most in this
section are
given in the following definition.

\proclaim {Definition 2.5} We let, for $\epsilon = \pm1 \text{ and }
m \in \Bbb Z \setminus \{ 0 \},$
$$W_m ^{\epsilon}(z) :=\Sigma_{k \in \Bbb Z}((e(k,m)+\epsilon q^{m/2}f(k+1,m))
z^{-(2k+1)}+(\frac{ ( \epsilon q^{m/2} -1)}{2} h(k,m)+
\frac{ ( \epsilon q^{m/2} +1)}{2} s^k t^m I)z^{-2k}).$$
We also let
$$W_0 ^{-1}(z) :=\Sigma_{k \in \Bbb Z}((e(k,0)-f(k+1,0))z^{-(2k+1)}
-h(k,0)z^{-2k})+(1/2) c_s,\text{ and }$$
$$\alpha (z) :=\Sigma_{ k \in \Bbb Z }((e(k,0)+f(k+1,0))
z^{-(2k+1)}+ \Sigma_{ k \in \Bbb Z \setminus \{ 0 \}} s^k I z^{-2k}).$$
\endproclaim
Here we have fixed, once and for all, a square root of $q$ so the two
square roots of $q^m \text{ are } \epsilon q^{m/2}, \epsilon = \pm1.$
Notice that as $q$ is generic, so not a root of unity, the entries
$\epsilon q^{m/2}, \epsilon q^{m/2} -1, \epsilon q^{m/2} +1,$
 in the above series are
non-zero. It easily follows from this that the linear span of the coefficients
of $z^n \text{ for } n \in\Bbb Z$, together with $c_s \text{ and } c_t,$
is all of $\LL$ and moreover that these coefficients are linearly
 independent. Sometimes these coefficients are called the moments of
the series in question, so we have that the moments of $ W_m ^{\epsilon}(z),
W_0 ^{-1}(z),\text{ and } \alpha(z),$ together with $c_s,c_t$ form a basis of
$\LL$.  Another way to write these series, using  the notation
developed in Definition 1.31 together with delta functions,
  which will be very useful for our purposes,
  is as follows.  The verification that these formulas hold is immediate.
$$\Wme(z)=\delta(z^2/s) \pmatrix \epsilon q^{m/2} & z^{-1} \\
\epsilon q^{m/2} z & 1 \endpmatrix t^m,\text{ for } \epsilon = \pm 1,
m \in \Bbb Z \setminus \{ 0 \}. \tag2.6$$
$$\W0-1(z)=\delta (z^2/s) \pmatrix -1 & z^{-1} \\ -z & 1 \endpmatrix
 +(1/2)c_s.\tag2.7$$
$$\a (z)= \Sigma_{k \in \Bbb Z \setminus \{ 0 \}} A_k z^{-2k} +
\Sigma_{k \in \Bbb Z} E_{2k+1}z^{-(2k+1)}.\tag2.8$$
Our first Lemma  shows how the basis elements of our Heisenberg algebra
$\hat H $ interact with the elements $\Wme (z)$ for all possible
values of $\epsilon \text{ and } m.$
\proclaim{Lemma 2.9}
\roster
\item
$[E_{2k+1}, \Wme (z)]=
z^{2k+1}(1-(\epsilon q^{m/2})^{2k+1}) \Wme (z), \text{ and }
$
\item
$
[A_n,\Wme (z)]=z^{2n}(1-q^{mn}) \Wme (z), \text{ for }
m \in \Bbb Z \setminus \{ 0 \}, \epsilon = \pm 1 \text{ and also for }
\epsilon= -1,\quad m=0.$
\endroster
\endproclaim
\demo{Proof} Writing $\Wme (z)$ as above we obtain that$$\align  [E_{2k+1},
\Wme (z)]=
E_{2k+1} \delta(z^2/s)  \pmatrix \epsilon q^{m/2} & z^{-1} \\
\epsilon q^{m/2} z & 1 \endpmatrix t^m -
\delta(z^2/s)  \pmatrix \epsilon q^{m/2} & z^{-1} \\
\epsilon q^{m/2} z & 1 \endpmatrix t^m E_{2k+1} + \\
tr((d_s E_{2k+1}) \delta (z^2/s)  \pmatrix \epsilon q^{m/2} & z^{-1} \\
\epsilon q^{m/2} z & 1 \endpmatrix t^m)c_s +
tr((d_t E_{2k+1}) \delta (z^2/s)  \pmatrix \epsilon q^{m/2} & z^{-1} \\
\epsilon q^{m/2} z & 1 \endpmatrix t^m)c_t, \endalign $$ a sum of four
terms. Putting in
that $E_{2k+1}=\pmatrix 0 & s^k \\s^{k+1} & 0 \endpmatrix$
 (see 1.3.1) gives us the first term is $$ \align & \delta  (z^2/s) \pmatrix
\epsilon s^k q^{m/2} z & s^k \\ \epsilon s^{k+1} q^{m/2} & s^{k+1} z^{-1}
\endpmatrix t^m =
\delta  (z^2/s) \pmatrix
\epsilon z^{2k+1} q^{m/2}  & z^{2k} \\ \epsilon z^{2k+2} q^{m/2} & z^{2k+1}
\endpmatrix t^m  = \\ &
z^{2k+1} \delta(z^2/s)  \pmatrix \epsilon q^{m/2} & z^{-1} \\
\epsilon q^{m/2} z & 1 \endpmatrix t^m .\endalign $$ Here we have used the
identity
$$f(s,z) \delta (z^2/s)=f(z^2,z) \delta (z^2 /s), \tag2.10$$
which holds for Laurent polynomials $f(s,z).$
As for the second term, we must bring $E_{2k+1} \text{ across } t^m$ so we
must use that $ts=qst.$ Thus we have $$ t^m E_{2k+1}= \pmatrix 0 & s^k q^{km} \\
s^{k+1} q^{(k+1)m} & 0 \endpmatrix t^m.$$  Therefore we have that the second
term is
$$\align & \delta (z^2 /s) \pmatrix \epsilon q^{m/2} & z^{-1} \\
\epsilon q^{m/2} z & 1 \endpmatrix
 \pmatrix 0 & s^k q^{km} \\
s^{k+1} q^{(k+1)m} & 0 \endpmatrix t^m = \\
 & \delta (z^2 /s) \pmatrix z^{2k+1}q^{(k+1)m} & \epsilon q^{m/2} z^{2k}
q^{km} \\
z^{2k+2}q^{(k+1)m} & \epsilon q^{m/2} z^{2k+1} q^{km} \endpmatrix t^m = \\
& \epsilon (q^{m/2} z)^{2k+1} \delta (z^2/s)
\pmatrix \epsilon q^{m/2} & z^{-1} \\
\epsilon q^{m/2} z & 1 \endpmatrix t^m =\\
& ( \epsilon q^{m/2} z)^{2k+1} \delta (z^2/s)
\pmatrix \epsilon q^{m/2} & z^{-1} \\
\epsilon q^{m/2} z & 1 \endpmatrix t^m,\text { as } \epsilon =\pm 1 \text {
and }
2k+1 \text{ is odd. } \endalign $$
Here we have used identity (2.10) again.

   To deal with the third term we recall that $\varepsilon (s^k t^m) =0$
whenever $m \neq 0$ so that this third term is
$$ \align & \delta_{m,0} tr( \pmatrix 0 & k s^k \\ (k+1) s^{k+1} & 0 \endpmatrix
 \delta (z^2/s) \pmatrix -1 & z^{-1} \\ -z & 1 \endpmatrix ) c_s = \\
 &\delta_{m,0} \varepsilon (\delta (z^2/s) (-k z^{2k+1} +(k+1) z^{2k+1}))c_s=
\delta_{m,0} \varepsilon (\delta (z^2 /s) z^{2k+1} ) c_s=\delta_{m,0}
z^{2k+1} c_s.
\endalign $$
Finally we note the fourth term is zero because we have $d_t (E_{2k+1}) =0.$

   Thus we now have that $$\align & [E_{2k+1}, \Wme (z)]=(z^{2k+1} -
(\epsilon q^{m/2} z)^{2k+1}) \delta (z^2/s) \pmatrix \epsilon q^{m/2} &
z^{-1} \\
\epsilon q^{m/2} z & 1 \endpmatrix t^m +\delta_{m,0}z^{2k+1} c_s= \\ &
z^{2k+1}(1-(\epsilon q^{m/2})^{2k+1}) \Wme (z), \endalign $$
if $m \neq 0,\epsilon = \pm 1.$  In the case when $\epsilon =-1 \text{ and }
m=0$ we get
$$2z^{2k+1} \W0-1 (z) =z^{2k+1} (1-(\epsilon q^{m/2})^{2k+1}) \W0-1 (z).$$
This proves (1).

   To prove (2) we have
$$\align &[A_n,\Wme (z)]=[\pmatrix s^n & 0 \\ 0 & s^n \endpmatrix,
\delta(z^2/s) \pmatrix \epsilon q^{m/2} & z^{-1} \\
\epsilon q^{m/2} z & 1 \endpmatrix t^m] =\\
&(s^nI) \delta(z^2/s) \pmatrix \epsilon q^{m/2} & z^{-1} \\
\epsilon q^{m/2} z & 1 \endpmatrix t^m -
 \delta(z^2/s) \pmatrix \epsilon q^{m/2} & z^{-1} \\
\epsilon q^{m/2} z & 1 \endpmatrix t^m (s^n I) + \\
& tr((d_s(s^n I))\delta(z^2/s) \pmatrix \epsilon q^{m/2} & z^{-1} \\
\epsilon q^{m/2} z & 1 \endpmatrix t^m )c_s +\\
&tr((d_t(s^n I))\delta(z^2/s) \pmatrix \epsilon q^{m/2} & z^{-1} \\
\epsilon q^{m/2} z & 1 \endpmatrix t^m )c_t.\endalign $$
Notice that the last term here is zero, as $d_t(s^n)=0.$  Thus the above becomes
$$ \align & (z^{2n} -z^{2n}q^{mn}) \delta (z^2/s)
\pmatrix \epsilon q^{m/2} & z^{-1} \\
\epsilon q^{m/2} z & 1 \endpmatrix t^m +
\delta_{m,0}tr(ns^n \delta (z^2/s) \pmatrix -1 & z^{-1} \\ -z & 1
 \endpmatrix) c_s= \\
& z^{2n}(1- q^{mn}) \Wme (z),\endalign $$
if $m \neq 0.$ This is also true if $ m=0 \text{ and } \epsilon =-1$ as then
the right hand side, as well as the left hand, side is zero.
 This completes the proof of (2) and hence our Lemma.
\qed
\enddemo

    We next want to obtain formulas which express the commutators of
 the various series $\Wme (z)$.  Of course, for this we need to use
different formal variables, so we will consider the product
$[\Wmeo (z_1),\Wmet (z_2)] \text{ where } z_1,z_2 $ are formal variables.
Here it is convenient to simplify the notation and define
$$Q_m ^{\epsilon}(z) := \pmatrix \epsilon q^{m/2} & z^{-1} \\
\epsilon q^{m/2} z & 1 \endpmatrix . \tag2.11 $$
Thus we have taking the product that
$$\align & Q_{m_1} ^{\epsilon _1} (z_1) Q_{m_2} ^{\epsilon _2} (z_2)=\\
 & \pmatrix \epsilon_1 \epsilon_2 q^{(m_1+m_2)/2}+ \epsilon_2 q^{m_2/2} z_1
^{-1} z_2 &
\epsilon_1 q^{m_1/2} z_2 ^{-1} + z_1 ^{-1} \\
 \epsilon_1 \epsilon_2 q^{(m_1+m_2)/2}z_1+ \epsilon_2 q^{m_2/2}  z_2 &
(\epsilon_1 q^{m_1/2} z_1 /z_2) +1 \endpmatrix. \tag2.12 \endalign$$

\proclaim{Lemma 2.13} If $m_1 +m_2 \neq 0 $ then we have
$$ [\Wmeo(z_1), \Wmet (z_2)]=W_{m_1+m_2} ^{\epsilon_1 \epsilon_2}(z_1)
\delta (z_2/\epsilon_1 z_1 q^{m_1/2})-
W_{m_1+m_2} ^{\epsilon_1 \epsilon_2}(z_2)
\delta (z_1/\epsilon_2 z_2 q^{m_2/2}).  $$
\endproclaim
\demo{Proof}We can express the commutator in question as a sum of four terms
as follows.
$$\align & [\Wmeo(z_1), \Wmet (z_2)]= \delta( z_1 ^2/s) Q_{m_1} ^{\epsilon
_1} (z_1)
t^{m_1} \delta (z_2 ^2/s) Q_{m_2} ^{\epsilon _2} (z_2) t^{m_2} - \\ &
\delta (z_2 ^2/s) Q_{m_2} ^{\epsilon _2} (z_2) t^{m_2}
\delta( z_1 ^2/s) Q_{m_1} ^{\epsilon _1} (z_1)
t^{m_1} + \\ &
tr((d_s(\Wmeo (z_1)) \Wmet (z_2)) c_s + tr((d_t(\Wmeo (z_1)) \Wmet (z_2)) c_t.
\endalign $$
Using the delta function identities
$$ t^m \delta (z^2/s)= \delta (z^2/q^m s) t^m, \tag2.14 $$
$$\delta (z_1 ^2/s) \delta( z_2 ^2/ q^{m_1} s)=
\delta (z_1 ^2/s) \delta( z_2 ^2/ q^{m_1} z_1 ^2),\text { and } \tag2.15 $$
$$\delta( z_2 ^2/ q^{m_1} z_1 ^2)= (1/2)(\delta (z_2/ \epsilon q^{m_1 /2} z_1)
+\delta (z_2/- \epsilon q^{m_1/2} z_1)), \text { for } \epsilon = \pm 1,
\tag2.16 $$ on the first term above we get
$$ (1/2)\delta (z_1 ^2/s)(\delta (z_2/ \epsilon_1 q^{m_1 /2} z_1)
Q_{m_1} ^{\epsilon_1} (z_1) Q_{m_2} ^{\epsilon_2}(z_2)+
\delta (z_2/- \epsilon_1 q^{m_1 /2} z_1)
Q_{m_1} ^{\epsilon_1} (z_1) Q_{m_2} ^{\epsilon_2}(z_2))t^{m_1+m_2}.$$
Using (2.12) and (2.10) we find that
$$\delta (z_2/ \epsilon_1 q^{m_1 /2} z_1)
Q_{m_1} ^{\epsilon_1} (z_1) Q_{m_2} ^{\epsilon_2}(z_2)=
2\delta (z_2/ \epsilon_1 q^{m_1 /2} z_1) Q_{m_1+m_2} ^{\epsilon_1 \epsilon_2}
(z_2), \text{ and }$$
$$\delta (z_2/- \epsilon_1 q^{m_1 /2} z_1) Q_{m_1} ^{\epsilon_1} (z_1) Q_{m_2}
 ^{\epsilon_2}(z_2)=0.$$
Thus, the first two terms in
our expansion of $[\Wmeo(z_1), \Wmet (z_2)]$ become
$$ \align & \delta (z_1 ^2/s) \delta (z_2/ \epsilon_1 q^{m_1 /2} z_1)
Q_{m_1+m_2} ^{\epsilon_1 \epsilon_2}
(z_1)t^{m_1+m_2}-
\delta (z_2 ^2/s) \delta (z_1/ \epsilon_2 q^{m_2 /2} z_2)
Q_{m_1+m_2} ^{\epsilon_1 \epsilon_2}
(z_2)t^{m_1+m_2} \\ & =
 \delta (z_2/ \epsilon_1 q^{m_1 /2} z_1) W_{m_1+m_2} ^{ \epsilon_1
\epsilon_2}(z_1)
- \delta (z_1/ \epsilon_2 q^{m_2 /2} z_2)W_{m_1+m_2} ^{ \epsilon_1
\epsilon_2}(z_2).
\endalign $$
Finally we note that $tr((d_s(\Wmeo (z_1)) \Wmet (z_2)) c_s=
tr((d_t(\Wmeo (z_1)) \Wmet (z_2)) c_t=0$ because of the hypothesis $m_1+m_2
\neq 0$. This is what we want.
\qed
\enddemo

   In our next Lemma we treat the case of products of the form
$[\Wme (z_1), W_{-m} ^{-\epsilon}(z_2)] \text{ for } m \neq 0.$ The proof begins
the same as in the last Lemma but here the analysis of the coefficients
of $c_s \text{ and } c_t$ is a lot more delicate.

\proclaim{Lemma 2.17}For $m \neq 0, \epsilon = \pm 1$ we have
 $$[\Wme (z_1), W_{-m} ^{-\epsilon}(z_2)] =\W0-1 (z_1)
\delta (z_2/\epsilon z_1 q^{m/2})-\W0-1 (z_2)
\delta(z_1/-\epsilon z_2 q^{-m/2}).$$
\endproclaim
\demo{Proof}Exactly as in Lemma 2.13 we obtain that
$$\align & [\Wme (z_1), W_{-m} ^{-\epsilon}(z_2)]=\delta (z_1 ^2/s) \delta
(z_2/\epsilon q^{m/2}z_1) Q_0 ^{-1}(z_1) -
\delta (z_2 ^2/s) \delta
(z_1/-\epsilon q^{-m/2}z_2) Q_0 ^{-1}(z_2)+ \\ &
tr(d_s(\delta (z_1 ^2/ s) Q_m ^{\epsilon} (z_1) t^m)
\delta (z_2 ^2/ s) Q_{-m} ^{-\epsilon} (z_2) t^{-m})c_s + \\ &
tr(d_t(\delta (z_1 ^2/ s) Q_m ^{\epsilon} (z_1) t^m)
\delta (z_2 ^2/ s) Q_{-m} ^{-\epsilon} (z_2) t^{-m})c_t. \endalign $$
Using that $\W0-1 (z) -(1/2)c_s=\delta(z^2/s) Q_0 ^{-1}(z)$ we obtain that
the first two terms in the expansion of $[\Wme (z_1), W_{-m} ^{-\epsilon}]$
become
$$\align & \delta (z_2/\epsilon q^{m/2} z_1) \W0-1 (z_1)-
\delta (z_1/-\epsilon q^{-m/2} z_2) \W0-1 (z_2) \\ &
-(1/2)\delta (z_2/\epsilon q^{m/2} z_1)c_s +
(1/2)\delta (z_1/-\epsilon q^{-m/2} z_2)c_s. \tag2.18 \endalign $$
We will see that the third term in our expansion of
$[\Wme (z_1), W_{-m} ^{-\epsilon}(z_2)]$ will cancel with
$-(1/2)\delta (z_2/\epsilon q^{m/2} z_1)c_s +
(1/2)\delta (z_1/-\epsilon q^{-m/2} z_2)c_s$ in (2.18). Towards this end we
have
that this third term is
$$\align & tr(d_s(\delta (z_1 ^2/ s) Q_m ^{\epsilon} (z_1) t^m)
\delta (z_2 ^2/ s) Q_{-m} ^{-\epsilon} (z_2) t^{-m})c_s= \\ &
tr(Q_m ^{\epsilon} (z_1)Q_{-m} ^{-\epsilon} (z_2)) \varepsilon ((
d_s(\delta (z_1 ^2/ s))\delta (z_2 ^2/q^m s))c_s= \\ &
tr(Q_m ^{\epsilon} (z_1)Q_{-m} ^{-\epsilon} (z_2)) \delta^{(1)}
(z_2 ^2/q^m z_1 ^2)c_s . \endalign $$
Here we have used the following delta function identity.
$$\varepsilon ((
d_s(\delta (z_1 ^2/ s))\delta (z_2 ^2/q^m s))=\delta^{(1)}
(z_2 ^2/q^m z_1 ^2). \tag2.19 $$
   For simplicity we let
$$\mu(z_1,z_2)=tr(Q_m ^{\epsilon} (z_1)Q_{-m} ^{-\epsilon} (z_2)).\tag 2.20 $$
Also, we will next use the following two delta function identities,where
as usual $f(z_1,z_2)$ denotes a Laurent polynomial and $a$ is any non-zero
 scalar.
$$\delta^{(1)}(z^2)=(1/4)(\delta^{(1)}(\epsilon z) +
\delta^{(1)}(-\epsilon z)),\text{ for } \epsilon = \pm 1. \tag2.21$$
$$\align & f(z_1,z_2) \delta^{(1)}(az_1/z_2)=
f(z_1,az_1)\delta^{(1)}(az_1/z_2)+(D_{z_2}f(z_1,z_2))\delta(az_1/z_2)= \\ &
f(a^{-1}
z_2,z_2)\delta^{(1)}(az_1/z_2)-(D_{z_1}f(z_1,z_2))\delta(az_1/z_2).
\tag2.22 \endalign $$
Here we are using the notation that $D_{z_1} \text{ and }D_{z_2}$
 denotes the degree derivation
with respect to the variable $z_1 \text { and } z_2$ respectively. That is
$$D_{z_i}=z_i \frac{d}{dz_i} \text{ for } i=1,2. \tag 2.23$$
Our third term now becomes
$$\align & (1/4)\mu (z_1,z_2) (\delta^{(1)} (z_2/\epsilon q^{m/2} z_1)+
\delta^{(1)}(z_2/ -\epsilon q^{m/2} z_1)) c_s = \\ &
((1/4)\mu (\epsilon q^{-m/2}z_2,z_2)\delta^{(1)} (z_2/\epsilon q^{m/2} z_1)+
(1/4)(D_{z_1} \mu (z_1,z_2)) \delta (z_2/\epsilon q^{m/2} z_1) ) c_s + \\ &
((1/4)\mu (-\epsilon q^{-m/2}z_2,z_2)\delta^{(1)} (-z_2/\epsilon q^{m/2} z_1)+
(1/4)(D_{z_1} \mu (z_1,z_2)) \delta (z_2/-\epsilon q^{m/2} z_1) ) c_s
\endalign $$
Using (2.12) we find that
$$\mu (z_1,z_2)= \epsilon (q^{m/2} z_1 /z_2 -q^{-m/2}z_2 /z_1). \tag2.24 $$
Thus, doing the necessary evaluations we get that,
$$ \align & \mu (\epsilon q^{-m/2} z_2, z_2)= \epsilon (\epsilon - \epsilon )
=0, \\ &
\mu (-\epsilon q^{-m/2} z_2, z_2)= \epsilon (-\epsilon - (-\epsilon) )
=0,\\ &
D_{z_1}( \mu (z_1,z_2))=\epsilon (q^{m/2} z_1 /z_2 +q^{-m /2}z_2 /z_1 ),\\ &
D_{z_1}( \mu (z_1,z_2)) \mid_{z_2 = \epsilon q^{m/2} z_1}=
\epsilon (q^{m/2} \epsilon q^{-m/2} + q^{-m/2} \epsilon q^{m/2})
= 2 \epsilon^2 =2, \\ &
D_{z_1}( \mu (z_1,z_2)) \mid_{z_2 = -\epsilon q^{m/2} z_1}=-2. \endalign $$
This now gives us that the third term in our expansion of
$ [\Wme (z_1), W_{-m} ^{-\epsilon}(z_2)]$ is
$$(1/2)\delta (z_2/\epsilon q^{m/2} z_1)c_s -
(1/2)\delta (z_1/-\epsilon q^{-m/2} z_2)c_s$$ as we claimed above.

   Finally we need to deal with the coefficient of $c_t$ which comes from the
fourth term above. This is
$$\align & tr(d_t(\delta (z_1 ^2/ s) Q_m ^{\epsilon} (z_1) t^m)
\delta (z_2 ^2/ s)Q_{-m} ^{-\epsilon} (z_2) t^{-m})c_t=\\ &
m( \varepsilon (\delta (z_1 ^2/s) \delta (z_2 ^2/q^m s))
tr(Q_m ^{\epsilon} (z_1)Q_{-m} ^{-\epsilon} (z_2))c_t= \\ &
m( \epsilon ((q^{m/2} z_1 /z_2)-(q^{-m/2}z_2 /z_1)) \varepsilon
(\delta(z_1 ^2/s) \delta (z_2 ^2/ q^m z_1 ^2))c_t. \endalign$$
Here we have used (2.24). This now becomes
$$\align & m( \epsilon ((q^{m/2} z_1 /z_2)-(q^{-m/2}z_2 /z_1))
\delta ( z_2 ^2 /q^m z_1 ^2) c_t= \\ &
m( \epsilon ((q^{m/2} z_1 /z_2)-(q^{-m/2}z_2 /z_1))(1/2)
(\delta (z_2 /\epsilon q^{m/2} z_1 )+\delta (z_2 /-\epsilon q^{m/2} z_1 )c_t.
\endalign $$
Here we have used formula (2.16) as well as the
following delta function identity.
$$\varepsilon( \delta(z_1 ^2/s) \delta(z_2 ^2/q^m z_1 ^2))=
\delta(z_2 ^2/q^m z_1 ^2).\tag2.25 $$
Finally we see that this is
$$(1/2)m \epsilon (q^{m/2}(\epsilon q^{-m/2})-q^{-m/2}(\epsilon q^{m/2}))c_t,$$
which is easily seen to be zero. Thus we now have that
 $$[\Wme (z_1), W_{-m} ^{-\epsilon}(z_2)] =\W0-1 (z_1)
\delta (z_2/\epsilon z_1 q^{m /2})-\W0-1 (z_2)
\delta(z_1/-\epsilon z_2 q^{-m /2})$$
which is what we want.
\qed
\enddemo

   The next Lemma deals with products of the form
$[\Wme (z_1), W_{-m} ^{\epsilon}(z_2)]$. Here we will allow the case of
$m=0 \text{ and } \epsilon =-1.$ As many of the arguments are similar to
the previous two
 Lemmas we will be brief.

\proclaim{Lemma 2.26}For $m \in \Bbb Z \setminus \{ 0 \} \text{ and } \epsilon =
 \pm 1$ we have that
$$[\Wme (z_1), W_{-m} ^{\epsilon}(z_2)]=
(\alpha (z_1)-\alpha (z_2)) \delta (z_2/ \epsilon q^{m/2} z_1)+
\delta^{(1)}( z_2/\epsilon q^{m/2} z_1)c_s +
2m\delta ( z_2/\epsilon q^{m/2} z_1)c_t.$$
\endproclaim
\demo{Proof}Expanding as in Lemmas 2.13 and 2.17 we get that
$$\align &  [\Wme (z_1), W_{-m} ^{\epsilon}(z_2)]=\\ &
\delta (z_2/\epsilon q^{m/2} z_1) \delta (z_1 ^2/s) Q_0 ^1 (z_1)-
\delta (z_1/\epsilon q^{-m/2} z_2) \delta (z_2 ^2/s) Q_0 ^1 (z_2)+ \\ &
(tr(Q_m ^{\epsilon}(z_1) Q_{-m} ^{\epsilon}(z_2)))\delta^{(1)}(z_2 ^2/
q^m z_1 ^2)c_s + \\ &
m \varepsilon(\delta (z_1 ^2/s) \delta (z_2 ^2/q^m s))
(tr(Q_m ^{\epsilon}(z_1) Q_{-m} ^{\epsilon}(z_2))c_t. \endalign $$
Now $$Q_0 ^1 (z) = \pmatrix 1 & z^{-1} \\ z & 1 \endpmatrix=I+z^{-1}
\pmatrix 0 & 1 \\ z^2 & 0 \endpmatrix,$$
so that recalling  $E_1=\pmatrix 0 & 1 \\ s & 0 \endpmatrix$ we obtain
that
$$\align & \delta (z_1 ^2/s) Q_0 ^1 (z_1)=\delta (z_1 ^2/s) (I+z_1^{-1}
\pmatrix 0 & 1 \\z_1 ^2 & 0 \endpmatrix)= \\ &
\delta (z_1 ^2/s) I +z_1 ^{-1}\delta ( z_1 ^2/s) E_1= \\ &
I+ \Sigma_{k \in \Bbb Z \setminus \{ 0 \}} A_k z^{-2k} +
\Sigma_{k \in \Bbb Z} E_{2k+1}z^{-(2k+1)}=\a (z_1)+I.\endalign $$
Note that here we are in the $gl_2(\cq)$ version of our spaces as in
Remark 1.30.  Obviously we have that
$$\delta (z_2/\epsilon q^{m/2} z_1)=\delta (z_1/\epsilon q^{-m/2} z_2),$$
so that in the first two terms in our expansion of
$ [\Wme (z_1), W_{-m} ^{\epsilon}(z_2)]$ the identity term above cancels
 and we get
$$(\alpha (z_1)-\alpha (z_2)) \delta (z_2/ \epsilon q^{m/2} z_1).$$

   As for the third term it is
$$\align & (tr(Q_m ^{\epsilon}(z_1) Q_{-m}
^{\epsilon}(z_2)))\delta^{(1)}(z_2 ^2/
q^m z_1 ^2)c_s= \\ &
(1/4)\nu (z_1,z_2)(\delta^{(1)}(z_2/\epsilon q^{m/2}z_1)+
\delta ^{(1)}(-z_2/\epsilon q^{m/2} z_1))c_s,\endalign $$
where we have used (2.21) and have let
$$\nu (z_1,z_2)=tr(Q_m ^{\epsilon}(z_1) Q_{-m} ^{\epsilon}(z_2)). \tag2.27 $$
Next we use (2.22) as well as the following delta function identity
$$\delta^{(1)}(z)=-\delta ^{(1)}(1/z),\tag2.28 $$
to get the third term which is
$$\align & (-1/4) \nu (z_1,z_2)\delta ^{(1)}(\epsilon q^{m/2} z_1/z_2)+
(1/4) \nu (z_1,z_2) \delta ^{(1)}( -\epsilon q^{m/2} z_1/z_2)= \\ &
(-1/4)(\nu (z_1, \epsilon q^{m/2} z_1)\delta ^{(1)}(\epsilon q^{m/2} z_1/z_2)
+(D_{z_2}\nu (z_1,z_2))\delta ^{(1)}(\epsilon q^{m/2} z_1/z_2)- \\ &
(1/4)\nu (z_1, -\epsilon q^{m/2} z_1)\delta ^{(1)}(-\epsilon q^{m/2} z_1/z_2)
+(D_{z_2}\nu(z_1,z_2))\delta ^{(1)}(-\epsilon q^{m/2} z_1/z_2). \endalign $$
It is easy to see that
$$\align &  \nu (z_1,z_2)=1+ \epsilon q^{-m/2} z_1^{-1} z_2 +\epsilon
q^{m/2} z_1
z_2 ^{-1} +1, \\ &
\nu (z_1, \epsilon q^{m/2} z_1) =4, \\ &
\nu (z_1, -\epsilon q^{m/2} z_1)=0, \\ &
D_{z_2} \nu (z_1,z_2)=\epsilon q^{-m/2} z_2 z_1 ^{-1} -\epsilon
q^{m/2} z_1 z_2 ^{-1}, \\ &
(D_{z_2} \nu (z_1,z_2))\delta (\epsilon q^{m/2}z_1/z_2)=0, \\ &
(D_{z_2} \nu (z_1,z_2))\delta (-\epsilon q^{m/2}z_1/z_2)=0.\endalign $$
Thus, the third term is
$$-\delta ^{(1)}(\epsilon q^{m/2} z_1/z_2)c_s=
\delta ^{(1)}( z_2/\epsilon q^{m/2} z_1)c_s.$$
Finally, using (2.25) and (2.16) we get that the fourth term is
$$\align & m \nu(z_1,z_2) \delta (z_2 ^2/q^m z_1 ^2)c_t= \\ &
m \nu(z_1,z_2) (1/2) (\delta (z_2/ \epsilon q^{m/2} z_1) +
\delta (-z_2/ \epsilon q^{m/2} z_1))c_t= \\ &
(m/2)(\nu (z_1, \epsilon q^{m/2} z_1)\delta (z_2/ \epsilon q^{m/2} z_1)+
\nu (z_1, -\epsilon q^{m/2} z_1)\delta (-z_2/ \epsilon q^{m/2} z_1)c_t= \\ &
2m\delta (z_2/ \epsilon q^{m/2} z_1)c_t, \endalign $$
which is what we want.
\qed
\enddemo

   For the convenience of the reader we now summarize our previous work
so that we have an easy reference in one place for these results.
\proclaim{Theorem 2.29}The following formulas hold.
$$
\align & (1)\qquad[E_{2k+1},E_{2j+1}]=(2k+1)\delta_{k+j+1,0} c_s,\\ &
(2)\qquad [A_m,A_n]=2m\delta_{m+n,0} c_s,\\ &
(3)\qquad [E_{2k+1},A_m]=0,\text{ for all } k,j \in \Bbb Z \text{ and }
m,n \in \Bbb Z \setminus \{ 0 \},\\ &
(4)\qquad\a (z)= \Sigma_{k \in \Bbb Z \setminus \{ 0 \}} A_k z^{-2k} +
\Sigma_{k \in \Bbb Z} E_{2k+1}z^{-(2k+1)},\\ &
(5)\qquad[E_{2k+1}, \Wme (z)]=
z^{2k+1}(1-(\epsilon q^{m/2})^{2k+1}) \Wme (z), \text{ and }\\ &
(6)\qquad[A_n,\Wme (z)]=z^{2n}(1-q^{mn}) \Wme (z),\\ &
\text{ for }k \in \Bbb Z \quad m,n \in \Bbb Z \setminus \{ 0 \}, \epsilon =
\pm 1,
 \text{ and also for }
\epsilon= -1,m=0, \\ &
(7)\qquad[\Wmeo(z_1), \Wmet (z_2)]=W_{m_1+m_2} ^{\epsilon_1 \epsilon_2}(z_1)
\delta (z_2/\epsilon_1 z_1 q^{m_1/2})-
W_{m_1+m_2} ^{\epsilon_1 \epsilon_2}(z_2)
\delta (z_1/\epsilon_2 z_2 q^{m_2/2}) \\ &  \text { for } m_1 +m_2 \neq 0, \\ &
(8)\qquad [\Wme (z_1), W_{-m} ^{-\epsilon}(z_2)] =\W0-1 (z_1)
\delta (z_2/\epsilon z_1 q^{m /2})-\W0-1 (z_2)
\delta(z_1/-\epsilon z_2 q^{-m /2}), \\ & \text { for } m \neq 0,
 \epsilon = \pm 1, \\&
(9)\qquad[\Wme (z_1), W_{-m} ^{\epsilon}(z_2)]=
(\alpha (z_1)-\alpha (z_2)) \delta (z_2/ \epsilon q^{m/2} z_1)+ \\
&\delta^{(1)}( z_2/\epsilon q^{m/2} z_1)c_s +
2m\delta ( z_2/\epsilon q^{m/2} z_1)c_t, \\ &
\text { for } m \in \Bbb Z \setminus \{ 0 \} \text{ and } \epsilon =
 \pm 1. \endalign $$
\endproclaim

   \proclaim{Remark 2.30} Since the moments of both generating functions
$\Wme (z)$ and $\a (z)$, together
with the central elements $c_s,c_t$ form a basis of $\LL$, we have that in
 order to define a representation of $\LL$ it is enough to find a
countably infinite dimensional Heisenberg algebra and a module for it,
together with formal series in $z$ having coefficients in the endomorphisms
of this module, which satisfy (1) thru (9) of Theorem 2.29 as well
as having that $c_s,c_t$ are central. This will be
the method we use in the next section.
\endproclaim

\subhead {Section 3. The Principal Representation for our Algebra}\endsubhead
In this section we will construct a representation for the algebra $\LL$.
The
method we use will be to first construct a module, $\F $, for a countably
infinite dimensional
Heisenberg Lie algebra isomorphic to $\hat H $.  This module $\F$ will be the
tensor product
of a standard irreducible Heisenberg module with central charge $1$ and Laurent
polynomials,$\Bbb C [v,v^{-1}]$, where the variable $v$ is introduced to keep
track of the grading relative to the variable $t$ in our algebra $\LL$. The
central
 element $c_s$ will act as the identity on $\F$ while the central
 element $c_t$ will act as zero.  We will then construct elements of the space
$End(\F)[[z,z^{-1}]]$ which we show satisfy the formulas of Theorem 2.29.

   To begin we let $\Cal S$ be the polynomial algebra in the variables
$x_1,x_2,x_3, \dots , \text{ over } \Bbb C$ so that
$$\Cal S=\Bbb C[x_1,x_2,x_3,\dots ].\tag3.1 $$
As is usual we let $x_k$ denote multiplication on $\Cal S$ by the
variable $x_k$ while we let $ \frac{\partial}{\partial x_k}$ denote
the partial derivative on $\Cal S$ corresponding to the variable $x_k$.
Denoting the identity operator on $\Cal S \text{ by } c_s$ (this will
eventually be the image of the element $c_s \text{ of } \LL$) we have
$$[\frac{\partial}{\partial x_n},mx_m]=\delta_{n,m}mc_s. \tag3.2$$
It follows from 1.33, 1.34, and 1.35, that the assignment
$$ \align & E_{2k+1} \mapsto -(2k+1)x_{-(2k+1)}, \text{ for } k<0, \\ &
E_{2k+1} \mapsto \frac{\partial}{\partial x_{2k+1}},\text{ if } k \geq 0, \\ &
A_n \mapsto -2nx_{-2n} \text{ for } n \leq -1,\tag3.3  \\ &
A_n \mapsto \frac{\partial}{\partial x_{2n}} \text{ for } n \geq 1, \\ &
c_s \mapsto c_s= \text{identity on } \Cal S,  \endalign $$
defines a faithful representation of our Heisenberg algebra $\hat H \text{ on }
\Cal S$. Next, we let
$$\F=\Bbb C [v,v^{-1}] \otimes \Cal S,\tag3.4$$
where $v$ is a variable and the tensor product is over $\Bbb C$.
As is usual we still let $x_n$ denote multiplication on $\F$ by
the variable $x_n$ so that actually we now have $x_n=
Id_{\Bbb C [v,v^{-1}]} \otimes x_n$. Similarly we will use the notation
 $\frac{\partial}{\partial x_n}=Id_{\Bbb C [v,v^{-1}]} \otimes
\frac{\partial}{\partial x_n}$ as well as $c_s=Id_{\F}$.

   We are now ready to define the analogues of our operators $\alpha (z),
\text{ and } \Wme(z)$ which here we will denote by $a(z) \text{ and }
X_m^{\epsilon}(z),$ respectively.

\proclaim{Definition 3.5}Let $\epsilon = \pm1 \text{ and } m \in \Bbb Z
\setminus
\{ 0\} \text{ or let } \epsilon =-1 \text{  and } m=0.$ We define
$$\align &(1)\quad b(m,\epsilon):=1/(\epsilon q^{-m/2}-1),\\
&(2)\quad \quad a(z):=\Sigma_{m \geq 1}(\frac{\partial}{\partial x_m}z^{-m}+
mx_mz^m), \text{ and }\\
&(3)\quad \Xme (z):=v^m b(m,\epsilon)exp(\Sigma_{k \geq 1}z^k(1-
(\epsilon q^{m/2})^k)x_k)exp(-\Sigma_{k \geq 1}\frac{z^{-k}(1-
(\epsilon q^{m/2})^{-k})}{k} \frac{\partial}{\partial x_k}).
\endalign $$
\endproclaim

  Several remarks are in order. First, we note that the term $v^m$ appearing
in (3) above is to be understood as multiplication by $v^m \otimes Id_{
\Cal S} \text{ on } \F$.  Also, that these terms give well-defined
 elements of the space $End(\F)[[z,z^{-1}]]$ is by now standard and follows
as in [FLM]. We will see later why we need to pick the scalars $b(m,\epsilon)$
as we do in (1) above.  For the time being we begin to establish the various
formulas satisfied by these elements.

\proclaim{Lemma 3.6}  With $m \text{ and } \epsilon$ as in Definition 3.5 and
any $n \geq 1$ we have
$$[\frac{\partial}{\partial x_n},\Xme(z)]=z^n(1-(\epsilon q^{m/2})^n)\Xme(z),
\text{ and} \tag 1 $$
$$[nx_n,\Xme(z)]=z^{-n}(1-(\epsilon q^{m/2})^{-n})\Xme(z).\tag 2 $$
\endproclaim
\demo{Proof}  Clearly we have that
$$[\frac{\partial}{\partial x_n},exp(z^n(1-(\epsilon q^{m/2})^n)x_n)]=
z^n(1-(\epsilon q^{m/2})^n) exp(z^n(1-(\epsilon q^{m/2})^n)x_n).$$
It follows from this that
$$[\frac{\partial}{\partial x_n},\Xme(z)]=z^n(1-(\epsilon q^{m/2})^n)\Xme(z).$$
Similarly we note that
$$\align & [nx_n,exp((\frac{-z^{-n}(1-(\epsilon q^{m/2})^{-n})}{n})
\frac{\partial}{\partial x_n})]= \\ & z^{-n}(1-(\epsilon q^{m/2})^{-n})
exp((\frac{-z^{-n}(1-(\epsilon q^{m/2})^{-n})}{n})
\frac{\partial}{\partial x_n}), \endalign $$
and (2) follows easily from this.
\qed
\enddemo
\proclaim{Remark 3.7} The representation in 3.3 clearly lifts to a
 representation, $\pi$  where
$$ \pi: \hat H \rightarrow End(\Cal F) \subset End(\Cal F)[[z,z^{-1}]].$$
Then we find that the previous result implies that (5) and (6) of Theorem 2.29
holds in the present setting. That is, we have
$$\align & [\pi (E_{2k+1}), \Xme (z)]=
z^{2k+1}(1-(\epsilon q^{m/2})^{2k+1}) \Xme (z), \text{ and }\\ &
[\pi (A_n),\Xme (z)]=z^{2n}(1-q^{mn}) \Xme (z),\\ &  \text{ for }k \in \Bbb Z
 \text{ and } m,n \in \Bbb Z \setminus \{ 0 \}, \epsilon = \pm 1,
 \text{ and also for }
\epsilon= -1,m=0. \endalign $$
\endproclaim

   We now go towards verifying that (7),(8), and (9) of Theorem 2.29
hold for the operators $\Xme(z)$. To begin with we introduce some notation.
We let, as above, $z_1,z_2$ be variables and then let
$$\tau(z_1,z_2,\omega_1,\omega_2)=\frac{(1-(z_2/z_1))(1-(\omega_2 /\omega_1))}
{(1-(z_2/ \omega_1))(1-(\omega_2/z_1))}. \tag3.8 $$
Here $\omega_1, \omega_2 $ are  taken to be rational functions of the variables
 $z_1,z_2$
such that the above expression makes sense.
\proclaim{Lemma 3.9}For $m_i \in \Bbb Z \setminus \{ 0 \}, \epsilon_i = \pm 1
\text{ and for } m_i=0, \epsilon_i=-1 \text{ where } i=1,2$ we have that
$$\align & exp(-\Sigma_{k \geq 1}z_1 ^{-k}
\frac{(1-(\epsilon_1 q^{m_1/2})^{-k})}{k}
\frac{\partial}{\partial x_k})
exp(\Sigma_{k \geq 1}z_2 ^k (1-(\epsilon_2 q^{m_2/2})^k)x_k)= \\ &
\tau(z_1,z_2,\omega_1,\omega_2)
exp(\Sigma_{k \geq 1}z_2 ^k (1-(\epsilon_2 q^{m_2/2})^k)x_k)
 exp(-\Sigma_{k \geq 1}z_1 ^{-k}
\frac{(1-(\epsilon_1 q^{m_1/2})^{-k})}{k}
\frac{\partial}{\partial x_k}), \endalign $$
where $\omega_i = \epsilon_i q^{m_i/2} z_i, \text{ for } i=1,2$.
\endproclaim
\demo{Proof} We have that the left hand side of the above is
$$exp(-\Sigma_{k \geq 1}\frac{z_1 ^{-k}- \omega_1
^{-k}}{k}\frac{\partial}{\partial x_k})
exp(\Sigma_{k \geq 1}(z_2 ^k -\omega _2 ^k)x_k).$$
Using the Campbell-Baker-Hausdorff formula together with
 the fact that in a Heisenberg Lie algebra the derived algebra
 equals the center of the algebra we have that this term is
$$\align & Pexp(\Sigma_{k \geq 1}(z_2 ^k -\omega _2 ^k)x_k)
exp(-\Sigma_{k \geq 1}\frac{z_1 ^{-k}- \omega_1 ^{-k}}{k}
\frac{\partial}{\partial x_k}) \quad \text{ where } \\ &
P=exp(-\Sigma_{k \geq 1}(\frac{z_1 ^{-k}- \omega_1 ^{-k}}{k})(z_2
^k-\omega_2 ^k)).\endalign$$
Thus we have that since $ln(1-z)=-\Sigma_{k \geq1} \frac{z^k}{k}$ then
$$\align & P=exp(-\Sigma_{k \geq 1}(\frac{(z_2/z_1)^k-(\omega_2 /z_1)^k-
(z_2/ \omega_1)^k+(\omega_2 /\omega_1)^k}{k})) \\ & =
exp(ln(1-(z_2/z_1)))exp(ln(1-(\omega_2/z_1))^{-1})
exp(ln(1-(z_2/\omega_1))^{-1})exp(ln(1-(\omega_2/\omega_1))) \\ & =
\frac{(1-(z_2/z_1))(1-(\omega_2 /\omega_1))}
{(1-(z_2/ \omega_1))(1-(\omega_2/z_1))}=\tau(z_1,z_2,\omega_1,\omega_2)
\endalign $$
which is what we want.
\qed
\enddemo
We now use this result to get our analogue of (7) and (8) of Theorem 2.29.

\proclaim{Lemma 3.10} Let $ \epsilon_1,\epsilon_2 \text{ and } m_1,m_2$ be as
usual but assume that either $m_1 +m_2 \neq 0 \text{ or that }
m_1 +m_2 =0 \text{ but } \epsilon_1 \epsilon_2 =-1$. Then we have that
$$\align & [\Xmeo(z_1),\Xmet(z_2)]= \\ &
X_{m_1+m_2}^{\epsilon_1 \epsilon_2}(z_1) \delta (z_2/ \epsilon_1 q^{m_1/2}z_1)
-X_{m_1+m_2}^{\epsilon_1 \epsilon_2}(z_2) \delta (z_1/ \epsilon_2 q^{m_2/2}z_2).
\endalign $$
\endproclaim
\demo{Proof}As in Lemma 3.9 we let $\omega_i = \epsilon_i
q^{m_i/2} z_i, \text{ for } i=1,2.$  Clearly  we have that
$$\align & [\Xmeo(z_1),\Xmet(z_2)]= \\ &
v^{m_1+m_2} b(m_1, \epsilon_1) b(m_2, \epsilon_2)
(\tau (z_1,z_2,\omega_1 ,\omega_2 )-\tau (z_2,z_1, \omega_2 ,\omega_1 ))E \\ &
\text{where } E=exp(\Sigma_{k \geq 1}(z_1 ^k -\omega _1 ^k +z_2 ^k -
\omega_2 ^k)
x_k)exp(-\Sigma_{k \geq 1}\frac{z_1 ^{-k} -\omega_1 ^{-k} +z_2 ^{-k}
-\omega_2 ^{-k}}{k})
\frac{\partial}{\partial x_k}).\endalign $$
We will use the following easily established delta function identity,
$$ \delta (z)= \frac{1}{1-z} + \frac{1}{1- z^{-1}} -1
. \tag3.11  $$
We let $x= z_2/z_1 \text{ and } \alpha_i=\epsilon_i q^{m_i/2} \text{ for }
i=1,2.$
Then
$$ \tau (z_1,z_2,\omega_1 ,\omega_2 )= \frac{(1-x)(1-(\alpha_2/ \alpha_1 )x)}
{(1-\alpha_1^{-1} x)(1-\alpha_2 x)} \text{ and }$$
$$\tau (z_2,z_1,\omega_2 ,\omega_1 )=\frac{(1-x^{-1})(1-(\alpha_1
/\alpha_2)x^{-1})}
{(1-\alpha_2 ^{-1} x^{-1})(1- \alpha_1 x^{-1})}.$$
It follows easily, using partial fractions, that we have
$$\tau (z_1,z_2,\omega_1 ,\omega_2 )=1+C[\frac{1}{1- \alpha_1^{-1}x} -
\frac{1}{1- \alpha_2 x} ], $$
$$\tau (z_2,z_1,\omega_2 ,\omega_1 )=1+C[\frac{1}{1- \alpha_2^{-1}x^{-1}} -
\frac{1}{1- \alpha_1 x^{-1}}] ,$$
where
$$C= \frac{(1- \alpha_1) (1- \alpha_2)}{(1- \alpha_1 \alpha_2 )}.$$
Thus, using (3.11)  we have that
$$ \align & \tau (z_1,z_2,\omega_1 ,\omega_2 )-\tau (z_2,z_1,\omega_2
,\omega_1 )= \\ &
C[\frac{1}{1- \alpha_1^{-1}x} - \frac{1}{1- \alpha_2 x} -\frac{1}{1-
\alpha_2^{-1}x^{-1}} + \frac{1}{1- \alpha_1 x^{-1}}] = \\ &
C[ \delta (\alpha_1^{-1} x) - \delta (\alpha_2^{-1} x^{-1} )] = \\ &
C[ \delta (z_2/w_1) -\delta (z_1 /w_2)]. \endalign $$

Again we use that $\omega_i = \epsilon_i
q^{m_i/2} z_i, \text{ for } i=1,2$ to obtain that
$$ \align & E\delta (\omega_2 /z_1) = \\ &
exp( \Sigma_{k \geq 1}(z_2 ^k-(\epsilon_1 q^{m_1/2}z_1)^k)x_k)
exp(-\Sigma_{k \geq 1}\frac{z_2 ^{-k}-(\epsilon_1 q^{m_1/2}z_1)^{-k}}{k}
\frac{\partial}{\partial x_k})\delta(\epsilon_2 q^{m_2/2}z_2/z_1)= \\ &
exp(\Sigma_{k \geq1}z_2 ^k (1- (\epsilon_1 \epsilon_2 q^{m_1+m_2
/2})^k)x_k) \times \\ &
\times exp(-\Sigma_{k \geq 1}\frac{z_2 ^{-k}(1-(\epsilon_1 \epsilon_2
q^{m_1+m_2 /2})^{-k})}
{k}\frac{\partial}{\partial x_k})\delta(\epsilon_2 q^{m_2/2}z_2/z_1).
\endalign $$
Similarly one also gets that
$$ \align & E\delta (\omega_1 /z_2) = \\ &
exp(\Sigma_{k \geq1}z_1 ^k (1- (\epsilon_1 \epsilon_2 q^{m_1+m_2
/2})^k)x_k) \times \\ &
\times exp(-\Sigma_{k \geq 1}\frac{z_1 ^{-k}(1-(\epsilon_1 \epsilon_2
q^{m_1+m_2/2})^{-k})}
{k}\frac{\partial}{\partial x_k})\delta(\epsilon_1 q^{m_1/2}z_1/z_2).
\endalign $$
Finally we note that (recall Definition3.5 (1))
$$ \align & C=\frac{(1- \alpha_1) (1- \alpha_2)}{(1- \alpha_1 \alpha_2 )} = \\ &
\frac{b(m_1 +m_2 ,\epsilon_1 \epsilon_2)}{b(m_1 ,\epsilon_1) b(m_2
,\epsilon_2)}. \endalign $$
Thus we obtain that
$$\align & [\Xmeo(z_1),\Xmet(z_2)]= \\ &
X_{m_1+m_2}^{\epsilon_1 \epsilon_2}(z_1) \delta (z_2/ \epsilon_1 q^{m_1/2}z_1)
-X_{m_1+m_2}^{\epsilon_1 \epsilon_2}(z_2) \delta (z_1/ \epsilon_2 q^{m_2/2}z_2).
\endalign $$
This is what we want.
\qed
\enddemo

\proclaim{Remark 3.12} It is in proving the previous Lemma that one is led to
the correct definition of the constants $b(m,\epsilon)$ which appear in
our operators $\Xme(z).$ Indeed, knowing Theorem 2.29 (7) and (8),
as well as the
computation in the last Lemma, shows that we must take
 $b(m,\epsilon):=1/(\epsilon q^{-m/2}-1)$ as we did in Definition 3.5.
\endproclaim

    We deal with the one remaining case in our next Lemma.

\proclaim{Lemma 3.13}For $m \in \Bbb Z \setminus \{ 0 \},\epsilon = \pm 1$
 we have that
$$\align & [\Xme(z_1),X_{-m} ^{\epsilon}(z_2)]= \\ &
(a(z_1)-a(z_2))\delta (z_2/\epsilon q^{m/2}z_1)+
\delta^{(1)} (z_2/\epsilon q^{m/2}z_1)c_s.\endalign $$
\endproclaim
\demo{Proof}We let $\omega _i =\epsilon q^{m/2} z_i \text{ for } i=1,2$ so
we have as above that
$$
\align & [\Xme(z_1),X_{-m} ^{\epsilon}(z_2)]= \\ &
b(m,\epsilon)b(-m,\epsilon)(\tau(z_1,z_2,\omega_1,\omega_2)-
\tau(z_2,z_1,\omega_2,\omega_1))E, \endalign
$$
where $E$ is as in the previous Lemma and $\tau$ is as in (3.8). Letting
$$R(z_1,z_2)=\tau(z_1,z_2,\omega_1,\omega_2)-
\tau(z_2,z_1,\omega_2,\omega_1),$$ we have that
$$R(z_1,z_2)=\frac{(1-(z_2/z_1))(1-(q^{-m}z_2/z_1))}{(1-(z_2/\epsilon
q^{m/2}z_1))^2}
-\frac{(1-(z_1/z_2))(1-(q^m z_1/z_2))}{(1-(\epsilon q^{m/2}z_1/z_2))^2}.$$
Next, we use the delta function identity
$$\frac{1}{(1-z)^2}=z^{-1}\delta^{(1)}(z)+\frac
{z^{-2}}{(1-z^{-1})^2}, \tag3.14 $$
with $z=z_2/\epsilon q^{m/2}z_1$ to get
$$\align & R(z_1,z_2)=(1-(z_2/z_1))(1-(q^{-m}z_2/z_1))\frac{\epsilon
q^{m/2} z_1}
{z_2} \delta^{(1)}(z_2/\epsilon q^{m/2}z_1)+ \\ &
(1-(z_2/z_1))(1-(q^{-m}z_2/z_1))\frac{q^m z_1 ^2}{z_2 ^2} \frac{1}
{(1-(\epsilon q^{m/2}z_1/z_2))^2}-
\frac{(1-(z_1/z_2))(1-(q^m z_1/z_2))}{(1-(\epsilon q^{m/2}z_1/z_2))^2} \\ &
=(1-(z_2/z_1))(1-(q^{-m}z_2/z_1))\frac{\epsilon q^{m/2}z_1}{z_2}
\delta^{(1)}(z_2/\epsilon q^{m/2}z_1)+ \\ &
\frac{((z_1/z_2)-1)((q^mz_1/z_2)-1)}{(1-(\epsilon q^{m/2}z_1/z_2))^2}-
\frac{(1-(z_1/z_2))(1-(q^m z_1/z_2))}{(1-(\epsilon q^{m/2}z_1/z_2))^2}= \\ &
(1-(z_2/z_1))(1-(q^{-m}z_2/z_1))\frac{\epsilon q^{m/2}z_1}{z_2}
\delta^{(1)}(z_2/\epsilon q^{m/2}z_1). \endalign $$
We now let
$$\align &  g_1(z_1,z_2)=b(m,\epsilon)b(-m,\epsilon)
(1-(z_2/z_1))(1-(q^{-m}z_2/z_1))\frac{\epsilon q^{m/2}z_1}{z_2},\\ &
g_2(z_1,z_2)=\Sigma_{k\geq 1}(z_1 ^k-\omega_1 ^k+z_2 ^k -\omega_2 ^k)x_k, \\ &
g_3(z_1,z_2)=-\Sigma_{k \geq 1}\frac{(z_1 ^{-k}-\omega_1 ^{-k} +z_2 ^{-k}
-\omega_2 ^{-k})}
{k} \frac{\partial}{\partial x_k}.\endalign $$
Thus we have that
$$\align & [\Xme(z_1),X_{-m} ^{\epsilon}(z_2)]= \\ &
g_1(z_1,z_2)exp(g_2(z_1,z_2))exp(g_3(z_1,z_2))
\delta^{(1)}(z_2/\epsilon q^{m/2}z_1). \endalign $$
>From (2.22) and (2.28) we get the following identity for any non-zero
scalar $b$.
$$f(z_1,z_2)\delta^{(1)}(z_2/bz_1)=f(z_1,bz_1)\delta^{(1)}(z_2/bz_1)-
(D_{z_2}f(z_1,z_2))\delta(z_2/bz_1) \tag3.15 $$
Using this identity on the above (with $b=\epsilon q^{m/2}$) we get that
$$\align & [\Xme(z_1),X_{-m} ^{\epsilon}(z_2)]= \\ &
g_1(z_1,\epsilon q^{m/2}z_1)exp(g_2(z_1,\epsilon q^{m/2}z_1))
exp(g_3(z_1,\epsilon q^{m/2}z_1))\delta^{(1)}(z_2/\epsilon q^{m/2}z_1) \\ &
-(D_{z_2}g_1(z_1,z_2))exp(g_2(z_1,z_2))exp(g_3(z_1,z_2))\delta
(z_2/\epsilon q^{m/2} z_1) \\ &
-g_1(z_1,z_2)(D_{z_2}g_2(z_1,z_2))
exp(g_2(z_1,z_2))exp(g_3(z_1,z_2))\delta (z_2/\epsilon q^{m/2} z_1) \\ &
-g_1(z_1,z_2)exp(g_2(z_1,z_2))(D_{z_2}g_3(z_1,z_2))
exp(g_3(z_1,z_2))\delta (z_2/\epsilon q^{m/2} z_1).\endalign $$
It is easy to see that
$$\align & g_1(z_1,\epsilon q^{m/2} z_1)=1, \\ &
g_i(z_1,\epsilon q^{m/2}z_1)=0 \text{ for } i=2,3,\\ &
D_{z_2}g_1(z_1,z_2) \mid_{z_2=\epsilon q^{m/2}z_1}=0, \\ &
(D_{z_2}g_2(z_1,z_2)) \delta(z_2/\epsilon q^{m/2}z_1)=
(\Sigma_{k \geq 1}k(z_2 ^k -z_1 ^k)x_k)\delta(z_2/\epsilon q^{m/2}z_1), \\ &
(D_{z_2}g_3(z_1,z_2))\delta(z_2/\epsilon q^{m/2}z_1)=
(\Sigma_{k \geq 1}(z_2 ^k-z_1 ^k)\frac{\partial}{\partial
x_k})\delta(z_2/\epsilon q^{m/2}z_1)
\endalign $$
Our result follows easily from this.
\qed
\enddemo

   Recalling Remark 2.30 as well as Definition 2.5
 we have that the Lemmas proved so far in this
 section imply that we have now established the following result.

\proclaim{Theorem 3.15} The space $\F=\Bbb C [v,v^{-1}] \otimes S$ is a module
for the Lie Algebra $\LL$ where the central element $c_s$ acts as the
identity of $\F$ and the central element $c_t$ acts as zero.
Let $\pi$ be the corresponding representation.  Then the action of $\LL$
on $\F$ extends the action of $\hat H$ on $\F$ given in (3.3) and
satisfies that
for $m \neq 0, \epsilon = \pm 1,k \in \Bbb Z$
$$\align & \pi (e(k,m)+\epsilon q^{m/2}f(k+1,m)) \text{ equals the coefficient
 of }z^{-(2k+1)} \text{ in } \Xme(z), \\ &
\pi (\frac{ ( \epsilon q^{m/2} -1)}{2} h(k,m)+
\frac{ ( \epsilon q^{m/2} +1)}{2} s^k t^m I)
\text{ equals the coefficient
 of }z^{-2k} \text{ in } \Xme(z) \endalign $$
while for $m=0 \text{ and } \epsilon =-1,k \in \Bbb Z$ we have
$$\align & \pi (e(k,0)-f(k+1,0)) \text{ equals the coefficient
 of }z^{-(2k+1)} \text{ in } X_0 ^{-1}(z) \text{ and } \\ &
\pi (-h(k,0)+\delta_{k,0}(1/2)Id_F)
\text{ equals the coefficient
 of }z^{-2k} \text{ in } X_0 ^{-1}(z).\endalign $$
\endproclaim

   We next go towards showing that the representation $\pi$ above can be
 lifted to a representation of the Lie algebra $\Le$.  Thus, recalling 1.23
we see we only need to define operators $\pi (d_s),\pi(d_t)$ and then check
that they satisfy the appropriate formulas.  However, it turns out that it is
more convenient to work with the element $\hat d_s$ which is defined
by
$$\hat d_s :=2d_s+(1/2)ad(\pmatrix 1 & 0 \\ 0 & -1 \endpmatrix) =
2d_s +(1/2) ad(H).\tag3.16 $$
Notice that we have
$$\align & [d_t,\Wme(z)]=m\Wme(z), \\ &
[d_t,\alpha(z)]=0, \\ &
[d_t,c_s]=[d_t,c_t]=0, \text{ and } \\ &
[d_t,\hat d_s]=0. \endalign $$
The element $d_t$ will be represented by the degree derivation with respect
to our variable $v$ on the module $\F$ so that we will let
$$\pi (d_t) := D_v :=  v \frac{\partial}{\partial v}.\tag3.17 $$
Then, as follows easily from Definition 3.5, we have that
$$\align & [D_v,\Xme(z)]=m\Xme(z), \\ &
[D_v, a(z)]=0, \\ &
[D_v,\pi (c_t)]=[D_v,0]=0, \\ &
[D_v,\pi (c_s)]= [D_v, Id_F]=0, \text{ and so } \\ &
[D_v, \pi (\hat d_s)]=0. \endalign $$
This shows that it is easy to represent the element $d_t$. To deal with
$\hat d_s$ is a little more complicated but follows along the usual lines.
This is dealt with in the proof of our next result.

\proclaim{Corollary 3.18}The representation $\pi \text{ of } \LL$ extends to
a representation, which we also denote by $\pi$, of the Lie algebra
$\Le$ satisfying
$$\align & \pi (d_t)=D_v \text{ and } \\ &
 \pi(\hat d_s)=
-\Sigma_{k\geq 1}kx_k \frac{\partial}{\partial x_k}.\endalign $$
\endproclaim
\demo{Proof}  For the case when $m \neq 0 \text{ and } \epsilon = \pm1,$
and also for the case when $m=0,\epsilon= -1$ we have that (since $c_s$
is central)
$$\align & [\hat d_s,\Wme (z)]=
[2d_s+(1/2)\pmatrix 1 & 0 \\ 0 & -1 \endpmatrix,\delta (z^2/s)
\pmatrix \epsilon q^{m/2} & z^{-1} \\ \epsilon q^{m/2}z & 1 \endpmatrix t^m]=
\\ &
2d_s(\delta (z^2/s))\pmatrix \epsilon q^{m/2} & z^{-1} \\ \epsilon q^{m/2}z
& 1 \endpmatrix t^m+
\\ & (1/2)\delta(z^2/s)(\pmatrix \epsilon q^{m/2} & z^{-1} \\ -\epsilon
q^{m/2}z & -1 \endpmatrix -
\pmatrix \epsilon q^{m/2} & -z^{-1} \\ \epsilon q^{m/2}z & -1 \endpmatrix ) t^m=
\\ &
2d_s(\delta (z^2/s))\pmatrix \epsilon q^{m/2} & z^{-1} \\
\epsilon q^{m/2}z & 1 \endpmatrix t^m+ \delta (z^2/s)
\pmatrix 0 & z^{-1} \\ -\epsilon q^{m/2} z & 0 \endpmatrix t^m. \endalign $$
It is  straightforward to see that
$$2d_s(\delta (z^2/s))=-D_z(\delta (z^2/s)), \tag3.19 $$
so using this in the above gives us that
$$\align & [\hat d_s,\Wme (z)]=(-D_z(\delta (z^2/s)))
\pmatrix \epsilon q^{m/2} & z^{-1} \\ \epsilon q^{m/2}z & 1 \endpmatrix
t^m+ \\ &
\delta (z^2/s)(-D_z(\pmatrix \epsilon q^{m/2} & z^{-1} \\ \epsilon q^{m/2}z
& 1 \endpmatrix)
 t^m)= \\ &
-D_z(\Wme(z)), \endalign $$
so that we have
$$[\hat d_s,\Wme (z)]=-D_z(\Wme (z)).\tag3.20 $$
In a similar way we also obtain that
$$[\hat d_s, \alpha (z)]=-D_z(\alpha (z)). \tag3.21 $$

We must show similar formulas hold for $\pi (\hat d_s)$. In doing this
we will make use of the following two formulas.
$$[z \frac{\partial }{\partial z}, z^m]=mz^m. \tag 3.22 $$
$$[z \frac{ \partial }{ \partial z},(\frac{ \partial }{ \partial z})^m]=
-m(\frac{ \partial }{ \partial z})^m. \tag3.23 $$
Using these one easily obtains that
$$\align & [kx_k \frac {\partial}{\partial x_k},
exp(z^k(1-(\epsilon q^{m/2})^k)x_k)
exp(\frac {(-z^{-k}(1-(\epsilon q^{m/2})^{-k}))}{k} \frac {\partial
}{\partial x_k})]=
\\ &
kx_k z^k (1-(\epsilon q^{m/2})^k x_k)
exp(z^k(1-(\epsilon q^{m/2})^k)x_k)
exp(\frac {(-z^{-k}(1-(\epsilon q^{m/2})^{-k}))}{k} \frac {\partial
}{\partial x_k})+
\\ &
exp(z^k(1-(\epsilon q^{m/2})^k)x_k)
exp(\frac {(-z^{-k}(1-(\epsilon q^{m/2})^{-k}))}{k} \frac {\partial }
{\partial x_k})
\, k( \frac{z^{-k}(1-(\epsilon q^{m/2})^{-k})}{k}\frac{\partial }{\partial
x_k})=
\\ &
D_z(exp(z^k(1-(\epsilon q^{m/2})^k)x_k)
exp(\frac {(-z^{-k}(1-(\epsilon q^{m/2})^{-k}))}{k} \frac {\partial }{
\partial x_k})). \endalign $$
It follows from this that we have
$$[\pi (\hat d_s), \Xme (z)]=-D_z(\Xme(z)). \tag3.24 $$
Similarly, one  easily gets that
$$[ \pi(\hat d_s), a(z)]=-D_z(a(z)). \tag3.25 $$
Looking at (3.20) and (3.21) we see this is exactly what we need.
\qed
\enddemo

   We close with several results concerning the structure of the module $\F$.
First, we let $\Cal F_m:=\Bbb C v^m \otimes \Cal S \text{ for all } m \in
\Bbb Z$.
Clearly then, we have that $\Cal F_m $ is an irreducible $\hat H$-module for
each $m \in \Bbb Z$ and we have that
$$ \Cal F=\oplus_{m \in \Bbb Z} \Cal F_m . \tag3.26$$
Viewing the vertex operator $X_{\pm 1}^{\epsilon}(z)$ as a map from
$\Cal F_m \text{ to } \Cal F_{m \pm 1}[[z,z^{-1}]]$ we find that we have
$$ \align &  X_{\pm 1}^{\epsilon}(z)(v^m \otimes 1) =
v^{m \pm 1} b(\pm 1 ,\epsilon) exp( \Sigma_{k \geq 1}z^k(1-(\epsilon
q^{\pm 1/2})^k)x_k)= \\ &
v^{m \pm 1} b( \pm 1, \epsilon )[1 + z(1- \epsilon q^{\pm 1/2})x_1 + \dots ]
\endalign $$
which shows that for any $m \in \Bbb Z \text{ the } \Le -\text{module }$
generated by any $\Cal F_m$ is all of $\F$. These observations make the
following
 result almost clear.

\proclaim{Proposition 3.27} $\Cal F$ is an irreducible module for the
Lie algebra $\Le$.
\endproclaim
\demo {Proof}If $f$ is a non-zero element we can write $f=\Sigma_{ m \in \Bbb Z}
f_m,$ where each element $f_m \text{ is in } \F_m.$
We call the number of non-zero terms
in this sum the length of $f$. The remarks proceeding the statement of the
 Proposition show that any element of length one generates all of $\F$ as an
$\Le -\text{module}$. On the other hand if we have a non-zero submodule $M$
of $\F$ and if $f$ is a non-zero element of $M$ whose component $f_m \ne 0$
then we can write
$$f=f_m +\Sigma_{n \in \Bbb Z , n \neq m}f_n$$
and so we obtain that if the length of $f$ is greater than one then the
element $mf-\pi (d_v)f$ is again in $M$ and is non-zero and has length
shorter that $f$.  Our result follows from this.
\qed
\enddemo

\proclaim{Remark 3.28}Let $\Cal H $ be the span of the elements $ H,c_s,c_t
,d_s,d_t$ so that $\Cal H $ is the five dimensional Cartan subalgebra
of the EALA $\Le$. It is straightforward to see that the module $\F$ is a direct
sum of finite dimensional $\Cal H$ weight spaces.
\endproclaim

\enddocument